\documentclass[preprint,aps,amsfonts,nofootinbib,floatfix,11pt]{revtex4}
\usepackage{epsf}              %
\usepackage{latexsym}           %
\usepackage{amscd}              %
\usepackage{amsmath}            
\usepackage{graphicx,subfigure} %
\usepackage{comment}            %
\input xy
\xyoption{all}

\newcommand{\beq}{\begin{equation}}
\newcommand{\eeq}{\end{equation}}
\newcommand{\bea}{\begin{eqnarray}}
\newcommand{\eea}{\end{eqnarray}}

\newcommand{\benn}{\begin{displaymath}}
\newcommand{\eenn}{\end{displaymath}}

\def\slashchar#1{\ensuremath{                   %
   \setbox0=\hbox{${}#1{}$}                     
   \dimen0=\wd0                                 
   \setbox1=\hbox{/} \dimen1=\wd1               
   \ifdim\dimen0>\dimen1                        
   \rlap{\hbox to \dimen0{\hfil/\hfil}}         
   {}#1{}                                       
   \else                                        
   \rlap{\hbox to \dimen1{\hfil${}#1{}$\hfil}}  
   /                                            
   \fi}}                                        %
\def\simge{
 \mathrel{\rlap{\raise 0.511ex
  \hbox{$>$}}{\lower 0.511ex \hbox{$\sim$}}}}
\def\simle{
 \mathrel{\rlap{\raise 0.511ex
  \hbox{$<$}}{\lower 0.511ex \hbox{$\sim$}}}}

\begin{document}
\preprint{DOE/ER/40762-425}
\title{Coulomb corrections in quasi-elastic scattering: tests of the effective-momentum approximation
}

\author{S.~J.~Wallace}
 \affiliation{Department of Physics, University of Maryland, College Park,
              MD 20742}

\author{J. A. Tjon}
\affiliation{Physics Department, University of Utrecht, 3508 TA
Utrecht, The Netherlands }
\affiliation{Department of Physics,
University of Maryland, College Park, MD 20742}


\date{\today}

\begin{abstract}
Coulomb corrections for quasi-elastic scattering of electrons by
nuclei are calculated using eikonal distorted waves. Corrections to
the lowest-order eikonal approximation are included in order to
obtain accurate results. Spin-dependent eikonal phase shifts are
evaluated and they yield very small corrections to the longitudinal
and transverse cross sections at electron energies of 500 MeV or
higher.  Because of this the Rosenbluth procedure is accurate for
separation of the longitudinal and transverse response functions.
The effective-momentum approximation is also found to be accurate
with regard to removal of the remaining Coulomb effects from the
distorted waves. Calculations are presented for electron scattering
from $^{208}Pb$ and $^{56}Fe$ nuclei at energies of 500 MeV and 800
MeV and momentum transfers $q$ = 550 MeV/c and 900 MeV/c.

\end{abstract}


\maketitle

\section{Introduction}
Measurements of cross sections for quasi-elastic electron scattering provide basic tests
of our understanding of nuclei.   Experiments have been performed at the MIT
Bates Laboratory ~\cite{Altemus83, Deady83, Hotta84, Deady86,
Blatchley86, Dytman88, Dow88, Yates93, Williamson97}, at the
Saclay Laboratory~\cite{Barreau83, Meziani84, Meziani85,
Marchand85, Zghiche94,Gueye99} and at SLAC~\cite{Baran88, Chen91,
Meziani92} in order to explore this reaction. A recent review
of quasi-elastic scattering provides a guide to the experimental
and theoretical results~\cite{Benhar08}.  Generally it is assumed
that the reaction is dominated by the exchange of one hard photon between
the electron and a nucleon in the nucleus. Many soft photons also
are exchanged and their effects are described by distorted waves based on the Coulomb potential
in the initial and final states of the electron. An important issue is
to account for the effects of the Coulomb interaction in a manner that
allows extraction of the nuclear response functions, $R_L$ and $R_T$,
which correspond to plane-wave matrix elements of the
longitudinal and transverse parts of the currents.
After the nucleon form factor is divided out, the longitudinal response
function at a fixed and sufficiently large value of the momentum transfer, q,
is expected to satisfy a Coulomb
sum rule, i.e., $\int d\omega S_L(q,\omega) \approx Z$, where Z is the number
of protons.  Such a sum rule should hold under general conditions for
a nonrelativistic description of nuclear wave functions and currents provided only that $q$ is
sufficiently large to make the effects of correlations small. Although
 the response functions are affected by the final-state interactions (FSI),
the sum rule should not be affected by them because it relies
on a sum over a complete set of final states of the nucleus.  When relativistic
effects in the nuclear current are taken into account, there are
minor relativistic corrections to the sum rule but it remains useful.  However, attempts
to verify it experimentally have produced puzzling results and controversy about
Coulomb corrections.

    In order to obtain the longitudinal response function, it is conventional
to perform a Rosenbluth separation after dividing the cross section
by the Mott cross section $\sigma_M$ and some kinematical factors,
\begin{eqnarray}
\frac{1}{\sigma_M}\frac{d\sigma}{d\Omega d\omega}~ \epsilon ~
\frac{q^4}{Q^4} = \epsilon ~ R_L^{expt}(q,\omega) + \frac{1}{2}
\Biggr(\frac{q^2}{Q^2}\Biggr) ~ \ R_T^{expt}(q,\omega)
\label{eq:RLexpt&RTexpt}
\end{eqnarray}
with $Q^2 = \omega^2-q^2$ and where $\sigma_M= 4\alpha^2 E_f^2
cos^2(\frac{1}{2}\theta_e)/Q^4$. The longitudinal virtual-photon
polarization is defined as
\begin{equation}
 \epsilon = \Biggr( 1 + \frac{2q^2}{Q^2} tan^2 \frac{\theta_e}{2} \Biggr) ^{-1}.
\end{equation}
At fixed $q$ and $\omega$, it varies from 0 to 1 as the electron scattering angle $\theta_e$
varies from 180 to 0 degrees.  Measurements of the cross
section at different values of $\epsilon$ for fixed values of $q$ and $\omega$ allow
a separation of the experimental response functions, $R_L^{expt}$ and $R_T^{expt}$.
The $R_L^{expt}$ that is extracted from experimental cross
section as in Eq.~(\ref{eq:RLexpt&RTexpt}) can differ from the desired $R_L$, which
is the plane-wave matrix element of the longitudinal current, for two
reasons.  One is
that Coulomb corrections associated with spin-dependent
effects in the electron wave functions can cause the
contributions of the longitudinal and transverse currents to
have different dependence on $\theta_e$ than is assumed in
the Rosenbluth separation.~\cite{Co87} The other is that
Coulomb distorted waves alter the response. These
Coulomb corrections should be removed
before the Coulomb sum rule can be evaluated.
In addition, nucleon form factors should be divided out of the
longitudinal response
function.
A standard approach to modeling the Coulomb corrections is to use
 the distorted wave Born approximation (DWBA). One solves the Dirac equation
 for the electron distorted waves in the presence of the Coulomb potential
 for both the initial and final states of the electron.~\cite{Jin9294,Udias93,Udias95,Kim01}
 When these distorted waves
 are combined with nuclear wave functions obtained from a model of nuclear structure,
 and appropriate current operators,
 cross sections may be calculated and compared with experimental results.
  The DWBA analysis involves extensive numerical calculations. Owing to
the infinite range of the Coulomb potential, partial-wave expansions converge
 very slowly, the more so as the energy increases.

A number of works have used
 the eikonal approximation in order to simplify the analysis at high energies.
 The use of the eikonal approximation also must be accompanied by
 inclusion of "focusing factors" that are not present in the
 eikonal wave functions.~\cite{Yennie65, Lenz71,Rosenfelder80}
Czyz and Gottfried~\cite{Czyz63} used the eikonal approximation to
analyze electron scattering but that work did not include focusing
factors. An analysis based on the Schrodinger equation showed that
the corrections to the eikonal approximation generally produce a
focusing factor in the wave function.~\cite{Baker72} Work by Giusti
{\it et al.} also is based on the eikonal
approximation~\cite{Giusti87,Giusti88} and some recents works have
combined the eikonal approximation with semi-classical focusing
factors in order to assess Coulomb corrections in quasi-elastic
scattering.~\cite{Aste04a,Aste04b}

 A very simple effective-momentum approximation (EMA) for treating
 the Coulomb corrections was developed by Rosenfelder\cite{Rosenfelder80}
and Triani {\it et al.}~\cite{Traini01,Traini88}. In the EMA
 the effects of the Coulomb potential are incorporated as
 shifts of the initial and final electron momentum values
that should be used in a plane-wave
Born approximation (PWIA) analysis. The shifted electron momenta are the effective
momenta.
They imply a corresponding shift of the photon momentum, $q \rightarrow q_{eff}$.

Interpretations of quasi-elastic data depend upon many experimental
 details and different experiments have produced significantly different values of the
 Coulomb sum rule.~\cite{Morgenstern01,Benhar08} In addition to possible experimental differences,
 there are theoretical differences in the analysis of
 the Coulomb corrections because the DWBA analysis based on partial
 waves has been used for some experiments and the EMA analysis has
 been used for others. Sometimes it is assumed that the nucleon
form factors can be pulled out of the matrix element and evaluated at the
momentum transfer of the electron.~\cite{Kim01}  At other times the form factors
are evaluated at the effective photon momentum, $q_{eff}$.

 In order to address questions about the theoretical differences
in the treatment of Coulomb distorted waves, we developed a
 systematic eikonal expansion in Ref.~\cite{Tjon&Wallace06} that
 provides more accurate eikonal wave functions for a DWBA analysis.
 The accuracy is good enough to eliminate concerns about use of the eikonal approximation at the energies
 of interest. Moreover the "focusing factors"
 arise naturally as part of the corrections
 to the eikonal approximation and the ad-hoc procedure of incorporating
 them is replaced by a systematic procedure. The eikonal expansion was found to converge rapidly
 at electron energies of interest. It has the advantage of providing insight
 into the nature of the Coulomb corrections because the focusing factors,
the eikonal phase shifts that determine the momentum shifts and spin-dependent effects
can be isolated for study.

 In order to assess the accuracy of the EMA
 using the eikonal wave functions, we used in Ref.~\cite{Tjon&Wallace06} a very simple
 model of the nuclear wave functions and we neglected the spin-dependent
 Coulomb corrections for simplicity. In order to compare full DWBA calculations
 of $R_L$ with the results based on the EMA, the
DWBA results were fit to the EMA formula,
\begin{equation}
R_L({\bf q},\omega)= A ~ R_L^{PWIA}({\bf q}_{eff},\omega),
\label{eq:EMAfit}
\end{equation}
where the effective photon momentum is given by
\begin{equation}
{\bf q}_{eff} = \hat{k}_i \Big[ k_i - \delta k \Big] - \hat{k}_f
\Big[ k_f - \delta k \Big]. \label{eq:q_eff}
\end{equation}
There are two parameters in our EMA fits: the momentum-shift $\delta k$ and the
overall normalization constant $A \approx 1$.
Usually the parameter $A$ is assumed to be unity when experimental
data are fit using Eq. (\ref{eq:EMAfit}).  In model calculations that assumption
can be checked because the PWIA response is known in the model.
A factor $f_{EMA}$ can be used to relate $\delta k$ to the
Coulomb potential at $r=0$, as follows,
       \begin{equation}
       f_{EMA} = \frac{\delta k}{V_c(0)}.
       \end{equation}
The factor $f_{EMA}$ is approximately the same for
different nuclei.
       We found momentum shifts in Fig. 7 of Ref.~\cite{Tjon&Wallace06} that
       correspond to $f_{EMA}(\omega)\approx 0.7$ near the
peak of the response function for 500 MeV $e^-$ scatttering.  Larger
       values of $f_{EMA}$ up to about 1 were found at the smallest and largest
$\omega$ values but the response is small at those points.
Although fits of the DWBA results can be made more precise by allowing $\delta k$
to depend on the energy loss, $\omega$, in this work we
use a constant shift $\delta k$. That yields reasonable
results and is simpler and thus preferable for the analysis of experimental data.

    If the momentum-shift $\delta k$
 and normalization constant $A$
   are determined theoretically for a given nucleus such that the
DWBA response is well described by the EMA fit of Eq.~(\ref{eq:EMAfit}),
then one may equate $R_L({\bf q},\omega)$ at fixed electron
beam energy, $E$, and fixed momentum transfer ${\bf q}$ to a constant $A$
times the PWIA response function evaluated at the effective
momentum transfer.  That would remove the Coulomb effects to a reasonable
approximation and allow the PWIA response to be extracted
from experimental data.  We expect similar Coulomb corrections
for a variety of nuclear models.~\cite{Aste08}
The goal is to remove them with minimal reliance on any nuclear model.
However, it must first be determined how well $R_L$ can be extracted from
experimental data.

The spin-dependence of the eikonal wave functions was omitted in our
previous paper, which left unanswered the question of
the accuracy with which the desired response functions
might be extracted from experimental cross sections.
That is the first issue addressed in this paper.
  In Sec. II we restate the essential results of the eikonal expansion
  for Dirac wave functions and focus on the spin-dependent eikonal corrections. These are shown
  to provide very small differences to quasi-elastic cross sections, i.e.,
  the helicity matrix elements of the electron current are very
  close to those based on the PWIA. The consequence is that
  when both initial and final electron energies are 200 MeV or more,
  the usual Rosenbluth separation provides an accurate separation of
  $R_L$ and $R_T$, well within the limits of experimental accuracy.
Although the Rosenbluth separation should be accurate, there remain
significant Coulomb effects within $R_L$. They can be treated
  with reasonable accuracy by use of the effective momentum approximation.

 In order to determine more realistic values of
the momentum shift, $\delta k$, shell model wave functions are used
to describe the nucleus in this work. Our calculations are
simplified by use of an approximation that is introduced by us in
Ref.~\cite{TjW08} and described in Section III, and which is denoted
EMAr. That approximation applies the effective momentum
approximation to the hard-photon propagator and form factors in
order to reduce the numerical evaluation to a three-dimensional
integration that provides a careful treatment of the full ${\bf
r}$-dependence of the Coulombic effects from the electron wave
functions.  In the EMAr analysis, the nuclear current is handled in
terms of a hadronic tensor, which can be extended to include the
neutron contributions to cross sections. A comparison of the full
DWBA and EMAr calculations for the 1s shell of $^{208}Pb$ shows
close agreement of the results.

Section IV presents numerical calculations for quasi-elastic
scattering by $^{208}Pb$ and $^{56}Fe$ using kinematics that are
relevant to a recent experiment at Jefferson Laboratory. The nuclear
model used is simple and a number of significant effects are omitted
from the calculations, such as final-state
interactions~\cite{Horikawa80}, correlations,~\cite{Benhar89,Sick07}
and pion and $\Delta$ production~\cite{vanOrden, Dekker}, but the
calculated cross sections are expected to be roughly similar to
experimental ones. The main goal is to determine suitable fitting
parameters for use in applying Eq.~(\ref{eq:EMAfit}) to experimental
data so as to determine $R_L^{PWIA}$. Conclusions are presented in
Section V.

 \section{Quasi-elastic response functions}

 Because electron energies of interest generally are much greater than
the electron mass, and the Coulomb potential and photon
exchange are vector interactions, electron helicity is conserved to
a very high degree of accuracy in quasi-elastic scattering.  For example, using 500 MeV electrons
one finds that the helicity is conserved except for terms of
relative order $m_e^2/E^2 \approx 10^{-6}$.  In this work we keep
only the effects that are consistent with helicity conservation.

The distorted-wave Born approximation is used with eikonal wave functions
for the electron.
For outgoing-wave (+) or incoming-wave (-) boundary conditions, the Dirac wave functions for potential $V(r)$
are written as~\cite{Tjon&Wallace06}
\begin{eqnarray}
\Psi^{(\pm)}_{k,\lambda}({\bf r}) &=& \begin{pmatrix} u^{(\pm)}({\bf r}) \cr 2 \lambda
u^{(\pm)}({\bf r})\end{pmatrix} \xi_{\lambda}, \nonumber \\
u^{(\pm)}({\bf r}) &=& \left( 1 - \frac{V}{E_2}\right) ^{1/2} e^{ik z
} e^{i \chi^{(\pm)}}e^{-\omega^{(\pm)}}e^{i \sigma_e \bar{\gamma}^{(\pm)}}
,
\label{eq:chi+}
\end{eqnarray}
where $\xi_{\lambda}$ is a two-component helicity spinor
and $\lambda = \pm \frac{1}{2}$ is the helicity eigenvalue.
The lower components of the Dirac spinor are simply $2\lambda $ times
the upper components because the electron mass is neglected.
The wave propagates in the $z$-direction, which is along the asymptotic momentum ${\bf k}$, and an impact vector
${\bf b}$ is defined as the part of ${\bf r}$ that is
perpendicular to the $\hat{z}$-direction.  The eikonal phases $\chi^{(\pm)}$, $\omega^{(\pm)}$
and $\bar{\gamma}^{(\pm)} = \gamma^{(\pm)} \pm i \delta^{(\pm)}$ are obtained from
integrals over the potential along the z-direction as
shown in Ref.~\cite{Tjon&Wallace06}.
The spin matrix in the
eikonal phase is $\sigma_e = \sigma \cdot  \hat{b} \times \hat{z}$,
the energy is $E$ and $E_2=E+m$.

\subsection{DWBA analysis}

The DWBA cross section for knockout of a nucleon of momentum ${\bf p}$
involves a two-dimensional integration over the angles of
the knocked-out nucleon as follows,
\begin{eqnarray}
\frac{1}{\sigma_M}\frac{d\sigma}{d\Omega d\omega}= \frac{Q^4}{cos^2(\frac{\theta_e}{2})}
\sum_{nlm} \int d\Omega_p \frac{p E_p}{(2\pi)^5}\bigr| {\cal M}_{nlm}\bigr|^2.
\end{eqnarray}
Omitting the final-state interactions of the nucleon,
the matrix element for quasi-elastic knockout involves a six-dimensional integration,
\begin{eqnarray}
{\cal M}_{nlm} &=& \frac{1}{(2\pi)^2}\int d^3q' \int d^3 r
\overline{\Psi}_{k_f,\lambda_f}^{(-)*}({\bf r})\gamma^{\mu}e^{-i
{\bf q'}\cdot{\bf r}} \Psi_{k_i,\lambda_i}^{(+)}({\bf r})
\frac{j_{N\mu}({\bf q'},{\bf p})}{{\bf q'}^2 - \omega^2} \psi_{nlm}({\bf q'}-{\bf p}) \nonumber \\
&=&  \frac{1}{(2\pi)^2}\int d^3q' \int d^3r e^{i({\bf q} - {\bf q'})\cdot {\bf r}}
 h_e^{\mu}({\bf r}) f_f f_i e^{ i \chi }\frac{j_{N\mu}({\bf q'},{\bf p})}{{\bf q'}^2 - \omega^2}\psi_{nlm}({\bf q'}-{\bf p})
\label{eq:M_DWBA}
\end{eqnarray}
where $j_{N\mu}({\bf q'},{\bf p})$ is the nucleon current, $h_e^{\mu}$
is a four-vector of helicity matrix elements of the electron current and
$\psi_{nlm}({\bf q'}-{\bf p})$ is the momentum-space wave function of a nucleon
in the nucleus with quantum numbers n (radial), l (angular momentum)
and m (z-component of angular momentum).  Subscripts $i$ and $f$ refer to the initial
and final electron states with asymptotic momenta ${\bf k}_i$
and ${\bf k}_f$ that provide the respective
z-directions for incoming and outgoing waves.
The exchanged photon has energy $\omega$ and momentum
${\bf q'}$, the initial electron helicity
is $\lambda_i$ and the final electron helicity is $\lambda_f$.
The momentum transferred by the electron is
${\bf q} = {\bf k}_i - {\bf k}_f$ and  it differs from
the momentum ${\bf q'}$ of the photon because of the Coulomb effects.
Nucleon form factors within the nucleon current depend on the
photon momentum, $q'$, that is integrated.
The sum of eikonal phases for incoming and
outgoing waves is $\chi = \chi_f^{(-)}({\bf r}) + \chi_i^{(+)}({\bf r})$ and
the focusing factors are defined by
\begin{eqnarray}
  f_i({\bf r}) &=& \Biggr( 1 - \frac{V}{E_{2i}}\Biggr)^{1/2}
  e^{-\omega_i^{(+)}}, \nonumber \\ f_f({\bf r}) &=&
  \Biggr( 1 - \frac{V}{E_{2f}}\Biggr)^{1/2} e^{-\omega_f^{(-)}},
  \label{eq:Dirac-focus}
  \end{eqnarray}
where $E_{2i} = E_{i}+m$ and $E_{2f} =E_f + m$.

\begin{figure}[h]
\includegraphics[width=10cm,bb=  0 0 612 792,clip]
{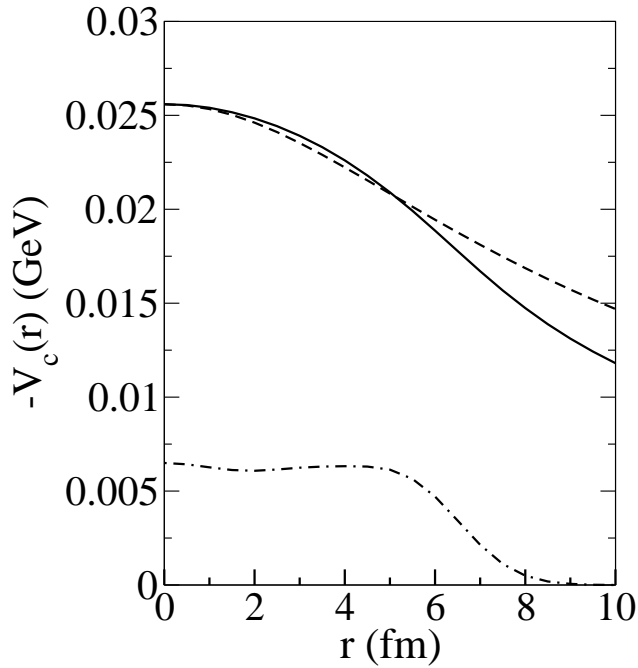} \caption{Coulomb potential for $^{208}Pb$.
The solid line shows the potential based on the empirically
determined charge density.~\cite{de_Vries87} The dash line shows the
approximate potential used in this work based on parameter values
$V_0$ = 0.0256 GeV and $R =$7.1 fm. The dash-dot line shows
one-tenth the charge density, $0.1 \rho_{ch}(r)$, for $^{208}Pb$.
}\label{fig:VCoulomb}
\end{figure}

\begin{figure}[h]
\includegraphics[width=10cm,bb=  0 0 612 792,clip]{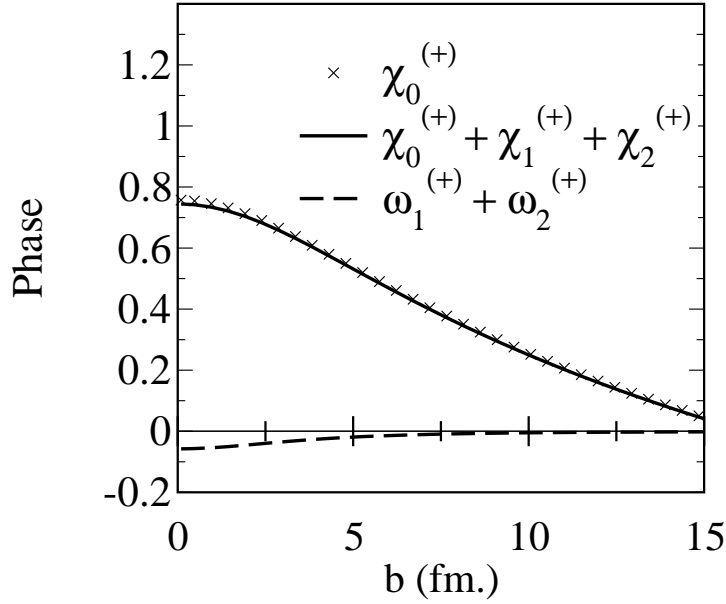}
\caption{ Eikonal phases at $z=0$ for 500 MeV electron scattering from
$^{208}Pb$.  }\label{fig:chi&omega}
\end{figure}

The Coulomb potential for scattering from a $^{208}Pb$ nucleus is shown in Fig.~\ref{fig:VCoulomb}.
The solid line shows the potential based on a fit of experimental
data for the charge density~\cite{de_Vries87} and
the dash line shows the simple potential used in our calculations, namely,
\begin{equation}
V_c(r) = \frac{-V_0}{(r^2+R^2)^{1/2}}.
\label{eq:VCoulomb}
\end{equation}
The simple potential allows an analytical evaluation of the eikonal
phases.  The parameters $V_0$ and $R$ are chosen so that the
simple potential has the same value as the empirical potential
at $r=0$ and the same average value as the potential based on experimental data
in the sense that
$\int dr \rho_{expt}(r) V_c(r) = \int dr \rho_{expt}(r) V_{expt}(r)$.  This ensures that
$V_c(r)$ provides a good fit in the range where the nuclear
density is significant (the dot-dash line shows one-tenth the nuclear charge density for reference).
Eikonal phases based on the Coulomb potential of Eq.~(\ref{eq:VCoulomb}) are shown
in Fig.~\ref{fig:chi&omega}.  The phases $\chi_i^{(+)}$ and $\omega_i^{(+)}$
are shown as a function of impact parameter $|{\bf b}|$
at $z=0$ and the nucleus is located at ${\bf b} = 0$, $z=0$.  Note that $\chi_i^{(+)}$ and $\chi_f^{(-)}$
each have the same behavior at $z=0$ and the total phase $\chi = \chi_i^{(+)}+\chi_f^{(-)}$
is approximately double the values shown.  Figure~\ref{fig:chi&omega}
shows that the eikonal expansion produces well converged results for the wave
function at electron energy equal to 500 MeV for scattering from $^{208}Pb$.
\vspace{0.25in}

Coulomb corrections affect the helicity matrix elements of the electron current
because of the spin-dependent eikonal phases involving $\bar{\gamma}^{(\pm)}$ as follows,
\begin{eqnarray}
h_e^{\mu}({\bf r})= \delta_{\lambda_f \lambda_i}
\xi^{\dag}_{\lambda_f}(\theta_e)
 e^{i\sigma_{e_f} \bar{\gamma}_f^{(-)*}}
\Big\{ 1, 2\lambda_i \vec{\sigma} \Big\} e^{i \sigma_{e_i}
\bar{\gamma}_i^{(+)}} \xi_{\lambda_i }.
\label{eq:ufbarOui}
\end{eqnarray}
 As shown in \cite{Tjon&Wallace06}, the required helicity matrix
 elements are given by
\begin{eqnarray}
h_e^0  &=&  A_{2 \lambda_i} cos{1 \over 2}\theta_e   + C_{2
\lambda_i} sin{1\over 2}\theta_e
\longrightarrow cos{1\over 2}\theta_e ,\nonumber \\
h_e^x  &=&   B_{2 \lambda_i} sin {1 \over 2}\theta_e  + D_{2
\lambda_i} cos{1 \over 2}\theta_e
\longrightarrow sin{1\over 2}\theta_e ,\nonumber \\
h_e^y&=&   2 i \lambda_i  \Bigr(A_{2 \lambda_i} sin{1 \over
2}\theta_e
  - C_{2 \lambda_i} cos{1 \over 2}\theta_e \Bigr)
\longrightarrow 2 i \lambda_i sin{1\over 2}\theta_e ,\nonumber \\
h_e^{z} &=&   B_{2 \lambda_i} cos{1 \over 2}\theta_e  -
D_{2 \lambda_i} sin {1 \over 2}\theta_e \longrightarrow cos{1\over 2}\theta_e .\nonumber \\
 \label{eq:helicityfacs}
\end{eqnarray}
The Rosenbluth separation implicitly assumes that the helicity
matrix elements take the plane-wave values that are shown following the
arrows in Eq.~(\ref{eq:helicityfacs}). One sees in Eq.~(\ref{eq:helicityfacs})
that the spin-dependent Coulomb effects cause each component of
the four-vector of electron helicity matrix elements to involve
both $cos{1 \over 2}\theta_e$ and $sin{1 \over 2}\theta_e$.
This property carries over to the longitudinal and transverse currents.
Consequently, $R_L^{expt}$ and $R_T^{expt}$ extracted using the Rosenbluth separation
each involve admixtures of the longitudinal and transverse currents.

\begin{figure}[h]
\includegraphics[width=10cm,bb=  0 0 612 792,clip]{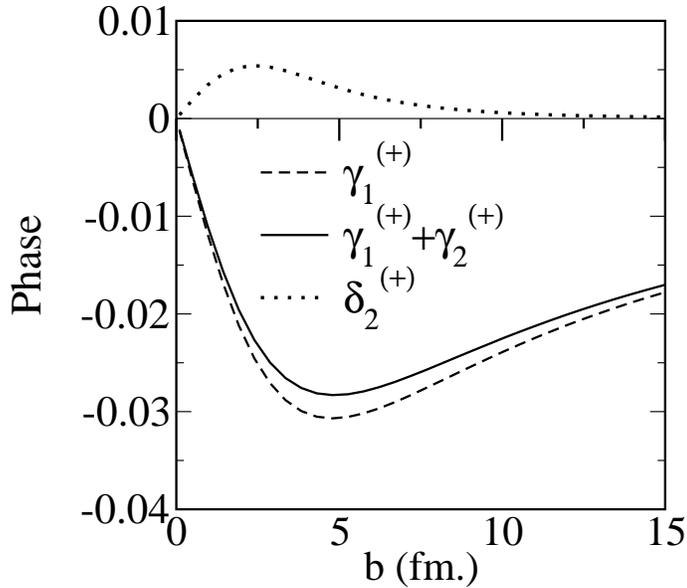}
\caption{ Spin phases for 500 MeV electron scattering from
$^{208}Pb$. }\label{fig:gamma&delta}
\end{figure}

The spin-orbit parts of the eikonal phases enter the
helicity matrix elements in the following four combinations
as shown in Ref.~\cite{Tjon&Wallace06},
\begin{eqnarray}
A_{2 \lambda_i}\equiv cos \bar{\gamma}_f^{(-)*} cos
\bar{\gamma}_i^{(+)} - sin\bar{\gamma}_i^{(+)}
sin\bar{\gamma}_f^{(-)*} e^{2i \lambda_i (\phi_i - \phi_f) }, \nonumber \\
B_{2 \lambda_i} \equiv  cos \bar{\gamma}_f^{(-)*} cos
\bar{\gamma}_i^{(+)}+
sin\bar{\gamma}_i^{(+)}sin\bar{\gamma}_f^{(-)*} e^{2i \lambda_i (\phi_i - \phi_f) }, \nonumber \\
C_{2 \lambda_i} \equiv cos
\bar{\gamma}_f^{(-)*}sin\bar{\gamma}_i^{(+)}e^{2i \lambda_i
\phi_i} + cos\bar{\gamma}_i^{(+)} sin\bar{\gamma}_f^{(-)*}
e^{-2i \lambda_i \phi_f },  \nonumber \\
D_{2 \lambda_i} \equiv  cos \bar{\gamma}_f^{(-)*}
sin\bar{\gamma}_i^{(+)}e^{2i \lambda_i \phi_i} - cos
\bar{\gamma}_i^{(+)} sin\bar{\gamma}_f^{(-)*} e^{-2i \lambda_i
\phi_f }. \nonumber \\  \label{eq:ABCD}
\end{eqnarray}
Figure~\ref{fig:gamma&delta} shows the eikonal phases $\gamma^{(+)}$ and $\delta^{(+)}$
that are the real and imaginary parts of $\bar{\gamma}^{(+)}$. There is a simple
relation between the phases as follows,
\begin{equation}
\bar{\gamma}^{(\pm)}({\bf r}) = \gamma^{(\pm)}({\bf r}) \pm i \delta^{(\pm)}({\bf r})
= \frac{1}{2(E+m)} \frac{d}{db} \Big( \chi^{(\pm)}({\bf r}) \pm i \omega^{(\pm)}({\bf r})\Big).
\end{equation}
This relation has been corrected from the one given in
Ref.~\cite{Tjon&Wallace06} because a factor $\frac{1}{2}$ was
omitted there.

\begin{figure}[h]
\includegraphics[width=10cm,bb=  0 0 612 792,clip]{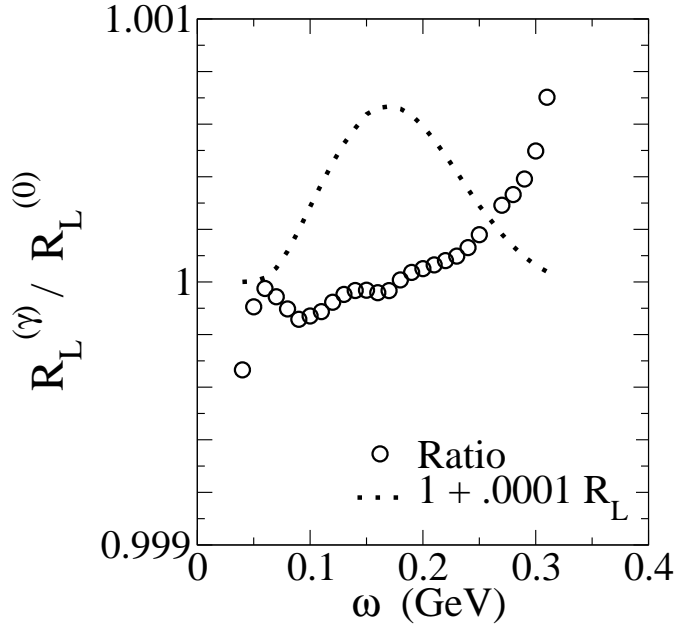}
\caption{ Circles show the ratio of the longitudinal response function $R_L^{(\gamma)}$ that includes
the spin-dependent Coulomb corrections to the longitudinal response
function $R_L^{(0)}$ that omits them for 500 MeV
electron scattering from $^{208}Pb$ at q=550 MeV/c. The dotted line shows
$1 + .0001R_L$. Note that the effects of the spin-dependent
Coulomb corrections are largest for $\omega$-values where the response function
is a small fraction of its maximum value.
}\label{fig:RL_EMAr&EMA}
\end{figure}

 Our previous numerical calculations omitted the
spin-dependent Coulomb corrections, using instead the PWIA
helicity matrix elements that are indicated following the arrows
in Eq.~(\ref{eq:helicityfacs}). In this work we include the
spin-dependent Coulomb corrections and find the important result
that they {\it provide negligible corrections} to quasi-elastic
cross sections.  This can be understood qualitatively as follows.
For a 500 MeV electron scattering from the Coulomb potential of a
$^{208}Pb$ nucleus, the phase $\bar{\gamma}^{(+)} = \gamma^{(+)} + i \delta^{(+)}$
typically has magnitude of 0.03 or less as shown in
Fig.~\ref{fig:gamma&delta}.
 It vanishes at zero impact
parameter.  Consequently, the coefficients $A_{2 \lambda_i}
\approx 1 + {\cal O}(\bar{\gamma})^2$ and $B_{2 \lambda_i}\approx
1 + {\cal O}(\bar{\gamma})^2$ are very close to unity. The
coefficients $C_{2 \lambda_i}$ and $D_{2 \lambda_i}$ have
contributions that are first order in $\bar{\gamma}$.  They also
involve phase factors $e^{2i \lambda_i \phi_i}$ or $e^{2i
\lambda_i \phi_f}$, where $tan \phi_f = tan \phi_i/\Big( cos
\frac{\theta_e}{2} - sin\frac{\theta_e}{2}/(tan \theta_i cos
\phi_i)\Big)$ with $\theta_i$ and $\phi_i$ being the angles of
the vector ${\bf r}$ relative to the polar axis along
$\hat{k}_i$. When current matrix elements are integrated over
${\bf r}$ and summed over helicities, cancellations stemming from these phase
factors cause the contributions to be
very small. Current matrix elements are Fourier transforms
and the $\phi_{i,f}$
dependent phase factors can receive support in the integration.
However, the quasi-free cross sections involve an additional
integration over angles of the knocked-out proton.
 The net effect is to reduce substantially the contributions from terms that involve the azimuthal
 angles $\phi_{i,f}$. Our numerical results show that
cross sections calculated with the spin-dependent Coulomb corrections
included are closer than one part per thousand to ones calculated using the PWIA helicity matrix elements.
An example of this is shown in Figure~\ref{fig:RL_EMAr&EMA}, which
shows the ratio of longitudinal response functions (calculated from Eq.~(\ref{eq:RL_EMAr}))
with and without the spin-dependent Coulomb corrections at 500 MeV electron energy.
Only for large energy loss, where the final state electron energy becomes small,
does the ratio differ from unity by more that a few parts in ten thousand.
 For reference, the dotted line shows the variation of
$1 + .0001 R_L$. Note that
for energy loss $\omega$ = 300 MeV, the final electron energy is 200 MeV, but the ratio of
cross sections with and without the Coulomb spin corrections differs from unity by less that
one part per thousand.   As the energy loss increases, the Coulomb spin corrections become
relatively more important but the response function is decreasing to zero.  The net effect
is that the absolute error in the response function is about one part in ten thousand of the maximum
value of $R_L$.
This is important because the helicity matrix elements govern the
dependence of the cross sections on the electron scattering
angle, $\theta_e$, at fixed $q$.
When the PWIA helicity matrix elements provide an accurate approximation, the angle dependence is
the same as in PWIA cross sections, and the Rosenbluth separation
can be used to extract the longitudinal and transverse current matrix elements.
We find negligible mixing of actual longitudinal and transverse current matrix elements in
the response functions extracted using the Rosenbluth separation for
all the cases evaluated in this paper.

Significant Coulombic effects remain in the $R_L$ and $R_T$ response
functions that can be extracted by use of the Rosenbluth separation
because of the spin-independent Coulomb effects.  They may be
treated using an effective momentum approximation.

\section{Effective photon momentum approximation}
Because the effective-momentum approximation was found
to be accurate in our prior work, in this work we approximate
the photon momentum ${\bf q'}$ that appears in the photon propagator
and the nucleon current of Eq.~(\ref{eq:M_DWBA}) by an effective momentum
${\bf q}_{eff}$ as follows,
\begin{eqnarray}
\frac{j_{\mu}^N ({\bf p}, {\bf q'})}{{\bf q'}^2 - \omega^2}  &&
\rightarrow \frac{j_{\mu}^N ({\bf
p}, {\bf q}_{eff})}{{\bf q}^2_{eff} - \omega^2},
\end{eqnarray}
where ${\bf q}_{eff}$ is given in Eq.~(\ref{eq:q_eff}).  Note that this approximation
also evaluates the nucleon form factors within the nucleon current
at the effective photon momentum.  This is a minimal use of
the effective-momentum approximation designed to reduce the
computation to a three-dimensional form, i.e.,
the approximation allows the photon propagator and the
nuclear current to be factored out of the integral over ${\bf q'}$, which
is then performed to obtain $\int \frac{d^3q'}{(2 \pi)^3}e^{-i {\bf q'} \cdot {\bf r}}\psi_{nlm}({\bf q'}-{\bf p})
= e^{-i {\bf p}\cdot {\bf r}} \psi_{nlm}({\bf r})$.
This yields
\begin{equation}
{\cal M}_{nlm} \longrightarrow {\cal M}^{EMAr}_{nlm} = \frac{2
\pi}{Q^2_{eff}}j_{\mu}^N({\bf p}, {\bf q}_{eff}) \,{\cal
M}^{\mu}_{nlm} \label{eq:EMAr}
\end{equation}
where $Q^2_{eff} = {\bf q}^2_{eff} - \omega^2$ and we define
\begin{equation}
{\cal M}^{\mu}_{nlm} = \int d^3r e^{i({\bf Q}- {\bf p})\cdot {\bf
r}}e^{i \chi({\bf r})}\, f_i({\bf r}) f_f({\bf r})\, h_e^{\mu}({\bf
r})\, \psi_{nlm}({\bf r}). \label{eq:M_mu}
\end{equation}
This procedure has been discussed in Ref~\cite{TjW08} and is called
the EMAr approximation. It has the advantage over the usual EMA of
not approximating the r-dependence of the eikonal phases and
focusing factors that provide the Coulomb corrections to the
electron wave functions. The three-dimensional integration of
Eq.~(\ref{eq:M_mu}) provides a good correspondence with the full
DWBA analysis at much lower computational cost. Procedures to
determine appropriate values of $\delta k$, and thus the effective
photon momentum that is factored out of the integral, are discussed
further on.


\subsection{Hadronic tensor}

   In the EMAr analysis, the bound-state nucleon's
wave function is taken to be a product of
a Dirac spinor, $u({\bf p}-{\bf q})$, times a nonrelativistic wave function for a
nucleon, i.e.,
$(M/(E_{{\bf p} - {\bf q}})^{1/2}u({\bf p}-{\bf q}) \psi_{nlm}({\bf p}-{\bf q})$,
and the knocked-out nucleon's wave function is a Dirac
spinor, $(M/E_p)^{1/2}u({\bf p})$. The relevant nucleon current is
\begin{equation}
j^N_{\mu} = K^{1/2} \bar{u}({\bf p}) \Big[ \gamma_{\mu} F_{1} +
\frac{i \kappa}{2M}F_{2} \sigma_{\mu \nu}q^{\nu} \Big] u({\bf p} -
{\bf q}),
\end{equation}
where $F_1(Q^2)$ and $F_2(Q^2)$ are nucleon form factors,
$\kappa$ is the anomalous magnetic moment and $
K =  M^2/(E_pE_{{\bf p}-{\bf q}})$
is a normalization factor arising from the spinors.
The EMAr cross section for knock-out of
a nucleon of momentum ${\bf p}$ by absorption of a photon
then takes the form
\begin{equation}
\frac{1}{\sigma_M} \frac{d\sigma}{d\Omega d\omega} = \frac{Q^4}{cos^2(\frac{\theta_e}{2})}
\sum_{nlm} \int d\Omega_p \frac{4
 p E_p}{(2\pi)^3}\frac{K}{(q^2_{eff}-\omega^2)^2}
{\cal M}^{\mu}_{nlm} W_{\mu \nu} (p, q_{eff}) {\cal M}^{\nu
\dag}_{nlm}
\end{equation}
where the hadronic tensor is,
\begin{eqnarray}
W_{\mu \nu}(p, q)= \frac{1}{2}Tr \Bigr[ \frac{p\!\!\!/ + M}{2M}
\Bigr( \gamma_{\mu}F_1 +i \frac{ \kappa}{2M}F_2 \sigma_{\mu \alpha}
q^{\alpha} \Bigr) \frac{p\!\!\!/ -q\!\!\!/ + M}{2M} \Bigr(
\gamma_{\nu}F_1 + i\frac{\kappa}{2M}F_2 \sigma_{\nu \beta} q^{\beta}
\Bigr) \Bigr] . \label{eq:W-alpha-beta}
\end{eqnarray}
Carrying out the trace over nucleon spins produces
\begin{eqnarray}
&&W_{\mu \nu} (p,q) =  \frac{1}{2} g_{\mu \nu}F_1^2 +
\frac{p_{\mu}(p-q)_{\nu}+ (p-q)_{\mu}p_{\nu} - p\cdot (p-q) g_{\mu
\nu} }{2M^2}F_1^2
\nonumber \\
&&+\Bigr(\frac{\kappa}{2M^2}F_1F_2+ \frac{\kappa^2}{8M^2}F_2^2\Bigr)
\Big(q^2 g_{\mu \nu} -q_{\mu} q_{\nu}\Big) \nonumber \\ &&+
\frac{\kappa^2}{8M^2}F_2^2 \Bigr[ -q_{\mu} q_{\nu} \big(p^2 + p\cdot
q\big) +2\big(q_{\mu} p_{\nu} + p_{\mu} q_{\nu}\big) \big( q^2 - p
\cdot q \big) + \big( 2 p \cdot q - q^2 \big) p \cdot q g_{\mu \nu}
\nonumber \\
&&-q^2 \big( 2 p_{\mu}p_{\nu} - p^2 g_{\mu \nu} \big)\Bigr].
\end{eqnarray}
The hadronic tensor is gauge invariant when the momenta are on
mass shell, i.e., $p^2 = M^2$ and $(p-q)^2 = M^2$.  These
conditions require that $p \cdot q = \frac{1}{2}q^2$.  We also use
$E_p = M+\omega$ for an initial nucleon at rest, leading to $K=M^2/[M(M+\omega)]$.
Using the on-mass-shell kinematics and $Q^2 = -q^2$ leads to the
gauge invariant form that is used in this work,
\begin{eqnarray}
W_{\mu \nu} (p, q) =  \frac{1}{M^2} \Bigr(p_{\mu} -
\frac{1}{2}q_{\mu}\Bigr) \Bigr( p_{\nu} -
\frac{1}{2}q_{\nu}\Bigr)\Bigr( F_1^2 + \frac{\kappa^2
Q^2}{4M^2}F_2^2\Bigr) + \frac{q^2 g_{\mu \nu} - q_{\mu} q_{\nu}
}{4M^2} \Big( F_1 + \kappa F_2\Big)^2 .
\end{eqnarray}
This form of the hadronic tensor is evaluated at $q^{\mu}
\rightarrow (\omega,{\bf q}_{eff)}$ in the EMAr analysis.

\subsection{Cross sections and response function}

The form of the hadronic tensor shows that the cross sections
involve an incoherent sum of two parts, which is a consequence of
averaging over nucleon spins.
The cross sections may be written concisely in terms of the Sachs form
factors,
\begin{eqnarray}
G_E &=& F_1-\frac{\kappa Q^2}{4M^2}F_2 \nonumber \\
G_M &=& F_1 + \kappa F_2
\end{eqnarray}
using the combination of form factors,
\begin{equation}
\widetilde{G}_E^2 = \frac{ G_E^2 + \frac{Q^2}{4M^2}G_M^2}{1 +
\frac{Q^2}{4M^2} },
\end{equation}
in place of $ F_1^2 + \frac{\kappa^2 Q^2}{4M^2}F_2^2$. Using current
conservation, which takes the effective forms $\vec{{\cal M}_{nlm}}
\cdot \hat{\bf q}_{eff} = (\omega/q_{eff}) {\cal M}^0_{nlm}$ and
${\bf h}_e\cdot \hat{q}_{eff} = (\omega /q_{eff}) h_e^0$, the
longitudinal components of the electron and nuclear currents can be
expressed in terms of the correspond charge components. In so doing
we arrive at
\begin{eqnarray}
\frac{1}{\sigma_M}\frac{d\sigma}{d\Omega d\omega} = \frac{Q^4}{cos^2(\frac{\theta_e}{2})}
\sum_{nlm}\int d\Omega_p \frac{ p E_p}{(2\pi)^3}\frac{K}{( Q^2_{eff})^2}
 \Biggr[ &&\Biggr|\frac{M+{1 \over 2}\omega}{M}{\cal M}^0_{nlm}
\Big( 1 - \frac{\omega^2}{{\bf q}^2_{eff}} \Big)  - \frac{{\bf p}}{M} \cdot \vec{{ \cal
M}}^T_{nlm}\Biggr|^2 \widetilde{G}_E^2(Q^2_{eff})  \nonumber \\
&&-{\cal M}_{nlm}^{\mu}g_{\mu \nu} {\cal M}_{nlm}^{\nu \dag}
\frac{Q_{eff}^2}{4M^2}G_M^2(Q^2_{eff}) \Biggr].
\label{eq:sigma_EMAr}
\end{eqnarray}
The transverse amplitude arising from the vector part
of the convection current is
\begin{equation}
\vec{{\cal M}}^T_{nlm} = \vec{{\cal M}}_{nlm} - \Big(\vec{{\cal M}}_{nlm} \cdot \hat{\bf q}_{eff}\Big) \hat{\bf q}_{eff}.
\end{equation}

As has been shown, it is a good approximation to omit
the spin-dependent eikonal effects.  Then
the helicity four-vector simplifies to the plane-wave form,
\begin{eqnarray}
h_e^{\mu PW}= \delta_{\lambda_f \lambda_i} \Big\{
cos\frac{\theta_e}{2}, sin \frac{\theta_e}{2},2\lambda_i sin
\frac{\theta_e}{2}, cos \frac{\theta_e}{2} \Big\}. \label{eq:he_PW}
\end{eqnarray}
It follows that
\begin{eqnarray}
|{\cal M}_{nlm}^0|^2 &&\propto cos^2\frac{\theta_e}{2},\nonumber \\
|\vec{{\cal M}}_{nlm}^T|^2 &&\propto sin^2\frac{\theta_e}{2}\nonumber \\
 {\cal M}_{nlm}^{\mu}g_{\mu\nu}{\cal M}_{nlm}^{\nu \dag}
&&\propto h_e^{\mu PW}g_{\mu\nu} h_e^{\nu PW} = -2 sin^2\frac{\theta_e}{2}.
\end{eqnarray}
The interference terms between ${\cal M}_{nlm}^0$ and
$\vec{\cal M}^T_{nlm}$ vanish by symmetry
and the quasi-elastic cross section takes the form of
Eq.~(\ref{eq:RLexpt&RTexpt})
with response function $R_L$ being proportional to the square of the matrix element of
the time-component of the current.
The longitudinal cross section is obtained as
\begin{eqnarray}
\frac{1}{\sigma_M}\frac{d\sigma_L}{d\Omega d\omega} =
 \frac{Q^4}{cos^2(\frac{\theta_e}{2})}
\sum_{nlm}\int d\Omega_p \frac{ p E_p}{(2\pi)^3}\frac{K}{( Q^2_{eff})^2}
 \Biggr[ \Biggr|\frac{M+{1 \over 2}\omega}{M}{\cal M}^0_{nlm}
\Big( 1 - \frac{\omega^2}{{\bf q}^2_{eff}} \Big)
\Biggr|^2 \widetilde{G}_E^2(Q^2_{eff}).
\end{eqnarray}

     This analysis suggests that to extract the longitudinal response
     function, one should perform the Rosenbluth separation as in Eq.~(\ref{eq:RLexpt&RTexpt})
     except that a factor ${\bf q}_{eff}^4/Q_{eff}^4$ should be used in place of the $q^4/Q^4$ kinematical factor in order
to cancel the $1 - \omega^2/{\bf q}_{eff}^2$ factor in the matrix element.
     Applying the factors ${\bf q}_{eff}^4/Q_{eff}^4$ and $1/\sigma_{M}$ to the longitudinal
     contribution of Eq.~(\ref{eq:sigma_EMAr}) yields
     \begin{equation}
     R_L^{EMAr} = \Biggr[\frac{(M+{1\over 2}\omega)^2}{M(M+\omega)}\Biggr] \Biggr[\frac{ Q^4}{Q^4_{eff}cos^2(\frac{\theta_e}{2})}\Biggr]
    \Biggr[\widetilde{G}^2_E(Q^2_{eff})\Biggr]
     \sum_{nlm} \int d\Omega_p \frac{pE_p}{(2\pi)^3} \Big|{\cal M}_{nlm}^0\Big|^2
     \label{eq:RL_EMAr} \end{equation}
     The first prefactor arises from the relativistic wave function normalization factor, K,
     and the current, where $p^0 = M+\omega$ and $q^0 = \omega$ have been used.
     The second prefactor should cancel to a large extent with similar
     factors in the ${\cal M}^0_{nlm}$ amplitude.  In particular, the helicity matrix element
     $h_e^0 \approx cos(\frac{\theta_e}{2})$ cancels to a high degree. To the extent that the focusing factors
     from the electron wave functions in the DWBA matrix element can be approximated
at ${\bf r}=0$, they give approximately a factor
     $\frac{k_i^{eff}k_f^{eff}}{k_ik_f}$ to the matrix element, which should cancel
  with $Q^4/Q^4_{eff} = \Big(\frac{k_ik_f}{k_i^{eff} k_f^{eff}}\Big)^2$ from the prefactor.
Because the EMAr analysis involves an integration over
${\bf r}$ with the full coordinate dependence,
the focusing factors can differ from the approximate result.
For example,
the contributions from radial wave functions that vanish at $r=0$, as in the shells with $L > 0$,
depend on the focusing factors away from the origin.
Consequently, accurate values of $\delta k$ are not known
a priori such that the second prefactor
cancels the Coulomb effects owing to focusing factors.
A procedure is needed to determine them.

   In order to test the Coulomb sum rule, one should remove
the nucleon form factor.
     Consistent with the photon momentum in the nuclear current being
shifted, the form factors evaluated at $Q^2_{eff}$ should be divided out
of the cross section, giving
\begin{eqnarray}
S_L(q,\omega) = \frac{1}{Z} \frac{R_L(q,\omega)}
{ \widetilde{G}^2_E(Q^2_{eff})}, \nonumber \\
\int d\omega S_L(q,\omega) \approx 1.
\label{eq:SL}
\end{eqnarray}

\subsection{Comparison of EMAr and DWBA analyses}

\begin{figure}[h]
\includegraphics[width=10cm,bb= 0 0 600 600 ,clip]{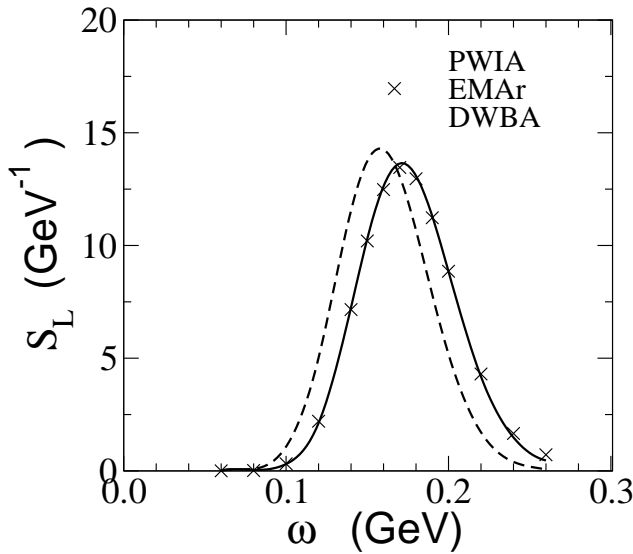}
\caption{ Response function for the 1s shell based on
the PWIA, EMAr and full 6D calculations
for $^{208}$Pb at E$_i$ = 500 Mev and
$q=550$ MeV/c.
}\label{fig:RL_PW_EMAr_6D.Pb}
\end{figure}

     The six-dimensional (6D) integration of Eq.~(\ref{eq:M_DWBA})
should be performed for nuclear models in order to obtain
the full DWBA response function.
Similar calculations of the EMAr response function should
then be normalized, by choice of the $Q_{eff}$ value
in the prefactor of the EMAr amplitude, such that the magnitude
of the peak DWBA response is reproduced
by the EMAr analysis.
     However, the full six-dimensional integration in the matrix element
together with the two-dimensional
integration over angles of the final momenta
     is extremely time consuming when many shells
     contribute to the response. In this work, a limited form of the
full DWBA analysis has been performed based on the wave function of
a single shell, the 1s shell of the $^{208}Pb$
nucleus. We then normalize the EMAr analysis to the DWBA
analysis for the 1s-shell.
 Figure~\ref{fig:RL_PW_EMAr_6D.Pb} shows the
PWIA, EMAr and DWBA response functions based on the 1s shell.
In the full DWBA calculation,
the nucleon form factor $F({\bf q'}^2-\omega^2)$, the photon propagator,
$1/({\bf q'}^2-\omega^2)$, and the current-conservation
factor $\Big(1 - \omega^2/{\bf q'}^2\Big)$ have been
evaluated at the
running photon momentum that is integrated as in Eq.(\ref{eq:M_DWBA}).
The response function is obtained by dividing
the resulting longitudinal cross section by the corresponding factors
evaluated at the effective photon momentum, $\widetilde{G}_E(q_{eff}^2-\omega^2)
\Big(1 - \omega^2 / q_{eff}^2\Big)/(q_{eff}^2-\omega^2)$.
In the EMAr analysis, there is an exact cancellation of
these factors.

The DWBA and EMAr results agree very well for the shift of the peak
response relative to the peak of the PWIA response function.  This is confirmed
by fits of each of these responses to the EMA form of Eq.~(\ref{eq:EMAfit})
using the parameters shown in Table~\ref{table:RL_EMAr_6D_fit}.
The shift of the 1s-shell peak response relative to the PWIA peak response is
fit by $\delta k$ of -22.0 MeV for the full DWBA case (labeled 6D-F$_{exact}$ in the table)
 and by
-22.0 MeV for the EMAr response function. The
magnitude of the peak EMAr response function is about 1.5\% lower than that for the DWBA
response function as shown by the A fit parameters in the table.

\begin{table}
\caption{EMA fit parameters for the 1s shell response functions at E=500 MeV and $q$=550 MeV/c.
$\delta$k is in MeV.
} \label{table:RL_EMAr_6D_fit}
\begin{tabular}{|cccc|}
\hline
   &  $~\delta$k~ &~~$f_{EMA}$~~&~~~A~~~  \\
\hline
$EMAr$~~  &  -22.0 &  0.80  &  0.985  \\
$6D-F_{exact} $   &  -22.0 &  0.80 &  1.00   \\
$6D-F_{eff} $ & -22.0  & 0.80   & 1.00   \\
\hline
\end{tabular}
\end{table}

In order to test one of the assumptions of the EMA or EMAr analysis,
we also calculated full DWBA cross sections with the form
factor and current-conservation factor
evaluated at $q_{eff}$ and removed from the integral over the
photon momentum. However, the photon propagator was left within
the integral. This allows a comparison of results for $S_L^{(DWBA)}$ with
the nonlocality of the form factor integrated over versus
otherwise identical results with the form factor evaluated at the effective momentum transfer
and taken out of the integral. In the latter case the form factor
simply cancels out of the result.
When $S_L^{(DWBA)}$ is calculated
with the form factor evaluated at $q_{eff}$ and factored out of the
integral, thus canceling, the result is fit using the parameters
in the line of Table~\ref{table:RL_EMAr_6D_fit} labeled 6D-F$_{eff}$.
When $S^{(DWBA)}_L$ is calculated with the form factor
integrated over and then divided out at $q_{eff}$, the result is fit
using the parameter values in the line
labeled 6D-F$_{exact}$ in the table.
Both calculations are found to produce essentially the
same response functions, $S_L^{DWBA}$, in the sense that the
same fitting parameters describe both equally well.
Thus, there is not any evidence
for errors associated with evaluating the form factor at $q_{eff}$
and factoring it out of the integral.
This is a nontrivial and important result because
the form factor reduces the cross sections by about a factor 4
for q=0.55 GeV/c.
There would be a significant difference
if the form factor evaluated at the momentum transfer of the
electron were divided out in Eq.(\ref{eq:SL})
as has been assumed to be the correct procedure in some works.
Use of the momentum transfer of the electron, q, versus the effective photon momentum, $q_{eff}$,
leads to a difference in cross sections by a factor
$\widetilde{G}_E^2(q^2-\omega^2)/\widetilde{G}_E^2(q_{eff}^2-\omega^2)\approx 1.23$
for $^{208}Pb$ at $q$ = 0.55 GeV/c and $\omega$ = 0.17 GeV.
We find very clear evidence from this analysis that
the form factor should be evaluated at the effective
photon momentum transfer rather than the momentum
transfer of the electron when response functions are
extracted from data.

We draw the following conclusions from these tests.  The EMAr approximation
provides a good approximation to the full 6D DWBA analysis.
It reproduces the shift of the DWBA response relative to the
PWIA response very well, i.e., the
momentum shift $\delta$k is the same: ($\delta$k$^{(6D}
= \delta$k$^{(EMAr)}$).
When the focus factors are kept within the integration over $r$ as in the
EMAr analysis, they
do not cancel precisely with the prefactor $Q^4/Q^4_{eff}$.
The $r$-integration provides a normalization reduction of about 1.5\%,
i.e., A = 0.985 in fits of the EMAr results to the EMA form.
In the 6D analysis with the nonlocality
of the photon propagator also included in the integration over photon momentum,
but everything else the same as in the EMAr calculation,
there is no normalization correction, i.e., A = 1.00 in fits to the
EMA form.  We conclude that the nonlocality of the photon
propagator produces a normalization
1.5\% greater than the normalization of the EMAr
response function: ( $R_L^{(6D)} \approx 1.015 R_L^{(EMAr)}$ ),
thus canceling the normalization reduction of the EMAr result.
In order to include the nonlocality of the photon propagator,
the normalization of the EMAr response for $^{208}Pb$ should be
increased by the factor 1.015.  It then agrees with the normalization
of the full DWBA result because the nonlocality of the
photon propagator cancels the reduction that arises in
the EMAr result.  The renormalized EMAr result
is found to give excellent agreement with the
full DWBA results for both the shift and the normalization.
When the nucleon form factor and
current-conservation factors also are kept in the
integration over photon momentum, there is
no additional change of the normalization
compared with evaluating those factors at $q_{eff}$.

%
%
%
\subsection{Comparison of EMAr and EMA calculations}

The three-dimensional integral of Eq.~(\ref{eq:M_mu})
is dominated by a stationary phase point that may be obtained by
approximating the eikonal phase $\chi({\bf r}) \approx \chi(0)
+ {\bf r}\cdot \nabla \chi(0) + \cdots$.  The effective
momentum is then
\begin{equation}
{\bf q}_{eff} = {\bf q} + \nabla \chi(0)
\end{equation}
and the integral for the time-component of the current takes the form
 [using $h_e^0 = cos\frac{1}{2}\theta_e$]
\begin{equation}
{\cal M}^{0,EMAr'}_{nlm} = cos\frac{1}{2}\theta_e
e^{i \chi(0)}\int d^3r e^{i({\bf q}_{eff}-
{\bf p})\cdot {\bf r}} f_i({\bf r}) f_f({\bf
r})  \psi_{nlm}({\bf r}).
\label{eq:M0_EMAr'}
\end{equation}
Generally it is found that the use of $q_{eff}$ overestimates the
Coulomb corrections unless $\nabla \chi(0)$ is reduced by
a factor in order to simulate an average value over the nucleus, i.e.,
\begin{equation}
{\bf q}_{eff} = {\bf q} + f_{EMA} \nabla \chi(0).
\end{equation}
We refer to this stationary-point analysis with the full
r-dependence of the focus factors left within the integral as EMAr'
and use $f_{EMA}$ = 0.8, as is consistent with fits of the
EMAr result.
When the focus factors are also approximated using
\begin{eqnarray}
f_i({\bf r})f_f({\bf r}) &\approx& \Big( 1 - f_{EMA} V(0)/k_i\Big) \Big( 1 - f_{EMA} V(0)/k_f\Big)
\nonumber \\
&=& \frac{k_{i, eff} k_{f,eff}}{k_i k_f}
\end{eqnarray}
then they are cancelled by the $1/Q_{eff}^2$ factor in the response
function.  That approximation leads to the usual EMA result,
\begin{eqnarray}
{\cal M}^{0,EMA}_{nlm} &&= cos\frac{1}{2}\theta_e
\frac{k_{i, eff} k_{f,eff}}{k_i k_f}
\psi_{nlm}({\bf q}_{eff} - {\bf p}).
\label{eq:M0_EMA}
\end{eqnarray}

\begin{figure}[h]
\includegraphics[width=10cm,bb= 0 0 600 600 ,clip]{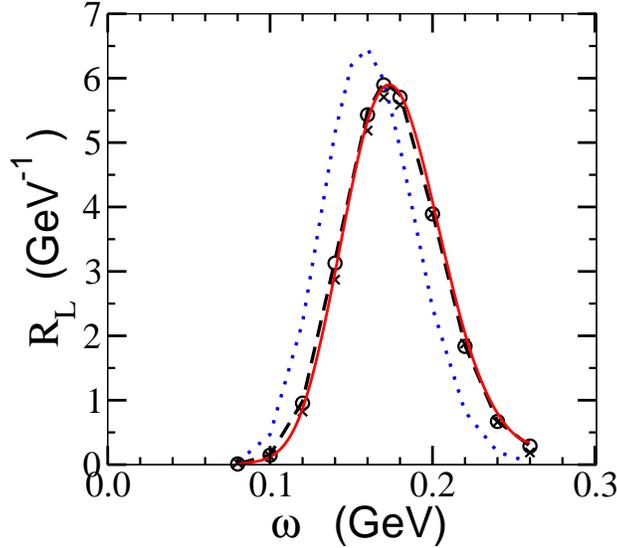}
\caption{ Longitudinal response function for the 1s-shell of $^{208}Pb$ based on
the PWIA (dotted line), EMA (o symbols), EMAr' (dash line), EMAr (solid line) and full 6D calculations (x symbols)
at E$_i$ = 500 Mev and
$q=550$ MeV/c.
}\label{fig:RL.e-.500.PW.EMA.EMAr.6D.Pb-1s.08.03.08}
\end{figure}

\begin{figure}[h]
\includegraphics[width=10cm,bb= 0 0 600 600 ,clip]{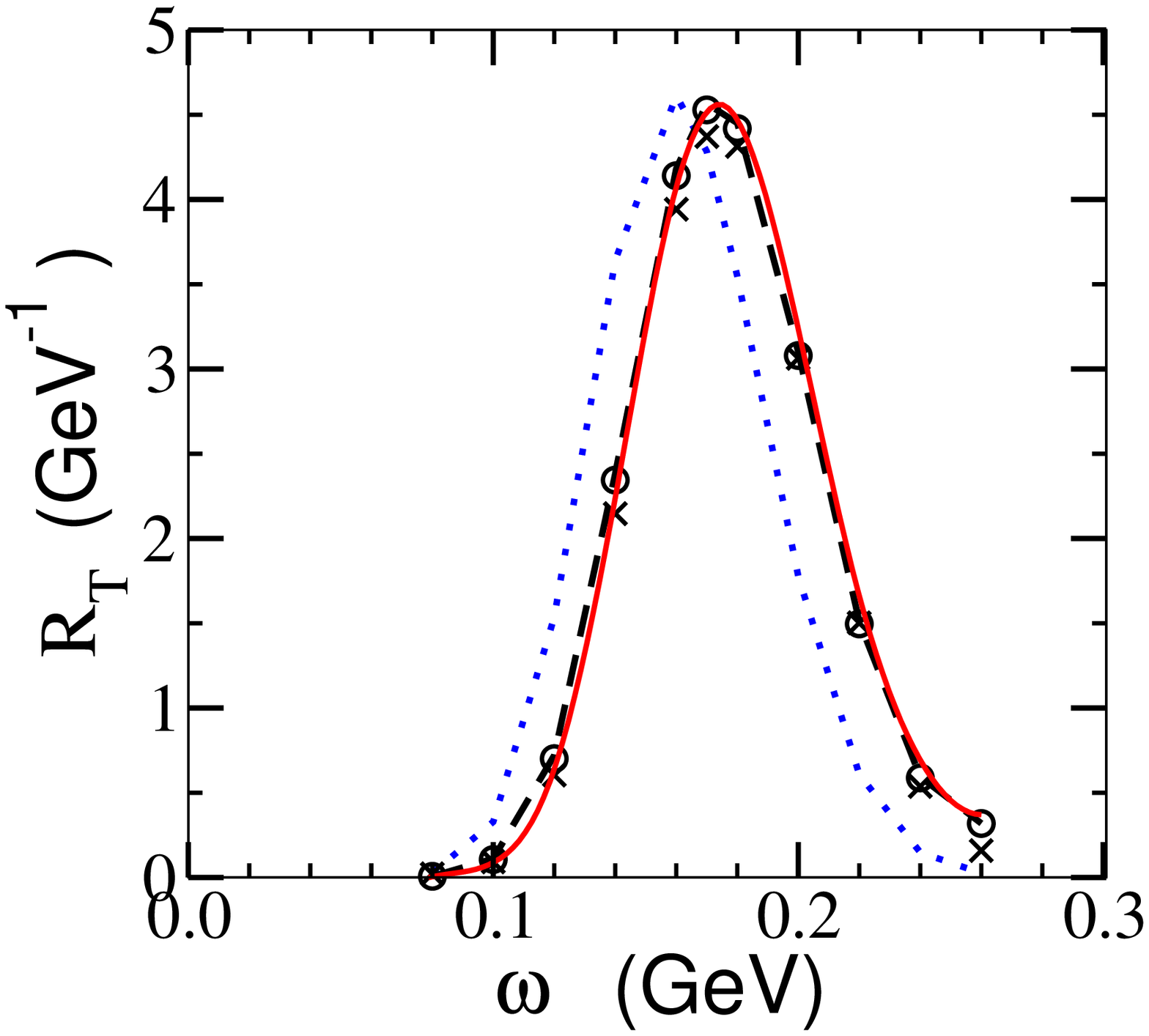}
\caption{ Transverse response function for the 1s-shell $^{208}Pb$ based on
the PWIA (dotted line), EMA (o symbols), EMAr' (dash line), EMAr (solid line) and full 6D calculations (x symbols)
at E$_i$ = 500 Mev and
$q=550$ MeV/c.
}\label{fig:RT.e-.500.PW.EMA.EMAr.6D.Pb-1s.08.03.08}
\end{figure}

The effective momentum approximations provide a good reproduction of 
the full $6D$ analysis for both the longitudinal response function, $R_L$, and the
transverse response function, $R_T$, 
as shown in Figures~\ref{fig:RL.e-.500.PW.EMA.EMAr.6D.Pb-1s.08.03.08}
and \ref{fig:RT.e-.500.PW.EMA.EMAr.6D.Pb-1s.08.03.08}.  
The EMAr' stationary-point analysis using $f_{EMA} = 0.8$
produces essentially the same results as the EMAr that
includes the integration over the variation of $\chi({\bf r})$.
The EMA result is also very close to the results based on EMAr and EMAr'.
Thus, it is clear that the integration over r that is
incorporated in the EMAr analysis provides results that differ only in the fine details.
The usual EMA analysis is almost as good once one has in hand a
reasonable value of $f_{EMA}$ to use.
Some numerical values are given in
Table~\ref{table:RL.RT.values} in order to provide a more quantitative
comparison of the approximations with the $6D$ calculation.  
 
\begin{table}
\caption{Numerical values for $R_L$ and $R_T$ for the 1s-shell of
$^{208}Pb$ at E=500 MeV and $q$=550 MeV/c.
} \label{table:RL.RT.values}
\begin{tabular}{|cccccc|}
\hline
  $\omega$  &  $R_L^{(PWIA)}$ &~~$R_L^{(EMA)}$ &  $R_L^{(EMAr)}$~~&~~~$R_L^{(EMAr')}$ & $R_L^{(6D)}$~~~ \\ 
\hline
  0.10  &   0.47 &  0.15 &  0.12 & 0.15 & 0.13 \\
  0.16  &   6.5  &  5.4  &  5.3  & 5.5  & 5.2 \\
  0.20  &   2.5  &  3.9  &  4.1  & 3.9  &  3.9  \\
\hline
  $\omega$  &  $R_T^{(PWIA)}$ &~~$R_T^{(EMA)}$ &  $R_T^{(EMAr)}$~~&~~~$R_T^{(EMAr')}$ & $R_T^{(6D)}$~~~  \\
\hline 
  0.10  & 0.33 &  0.11 & 0.09 & 0.11 & 0.09  \\ 
  0.16  & 4.6  &  4.1  & 4.1  &  4.2  & 3.9 \\
  0.20  &  1.9  &  3.1 &  3.2 &  3.1 & 3.1 \\
\hline
\end{tabular}
\end{table}

%
%

\section{Cross section calculations}

Numerical calculations using the EMAr analysis have been performed including all shells
of shell-model wave functions
with the harmonic oscillator parameter adjusted so that the correct nuclear
charge radius is obtained.  Table~\ref{tab:HO&V0params} shows
the parameter values used.  Harmonic oscillator
wave functions are used for the shell model.
In coordinate space they are
\begin{equation}
\psi_{nlm}({\bf r}) = N Y_{lm}(\Omega_r) r^l
\,_1F_1(-(n-l)/2,l+3/2,r\sqrt2/\beta)\, e^{-(r/\beta)^2}
\end{equation}
with normalization constants $N$ determined by  $\int d^3r
|\psi({\bf r})|^2 = 1$. Furthermore, $Y_{lm}$ are the well known
spherical harmonics and $_1F_1$ the confluent hypergeometric
functions.

\begin{table}
\caption{Parameters used in calculations: $\beta$ is the harmonic oscillator
parameter; $V_0$ and $R$ are the Coulomb
potential parameters.
} \label{tab:HO&V0params}
\begin{tabular}{|cccc|}
\hline
 Nucleus  &  $~\beta$(fm)~ &~~$V_0$ (GeV) ~~&~~~$R$~(fm)~~~  \\
\hline
$^{208}Pb$~&~ 3.564  & 0.0256 &  7.10  \\
$^{56}Fe $ &  2.854  & 0.0124 &  3.97  \\
\hline
\end{tabular}
\end{table}

Neutron contributions to cross sections are required
in order to include the magnetic scattering.  They are assumed to be proportional
to the proton contributions and are included by using suitable form factors, i.e.,
\begin{eqnarray}
\widetilde{G}_E^2 \longrightarrow \widetilde{G}_{Ep}^2+ \frac{N}{Z}\widetilde{G}_{En}^2,
\nonumber \\
G_M^2 \longrightarrow G_{Mp}^2+ \frac{N}{Z}G_{Mn}^2
\end{eqnarray}
times the proton contributions,
where subscripts $p$ and $n$ refer to the proton and neutron, respectively.
 Dipole form factors $1/(1 + Q^2/0.71GeV^2)^2$ are used for
the variation of $F_1$ and $F_2$ with $Q^2$.

\begin{figure}[h]
\includegraphics[width=7cm,bb=  4 15 553 533,clip]{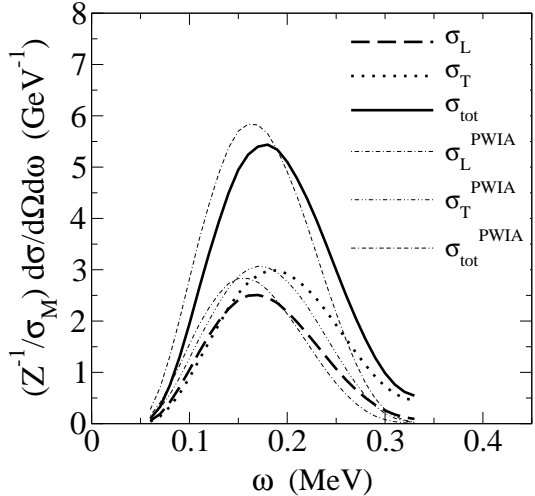}
\caption{ Ratio of EMAr and PWIA longitudinal, transverse and total cross
sections to Z times the Mott cross section for 500 MeV electron
scattering from $^{208}Pb$ at q=550 MeV/c.
}\label{fig:sigma.RUN1.Pb}
\end{figure}

\begin{figure}[h]
\includegraphics[width=10cm,bb= 0 0 612 792,clip]{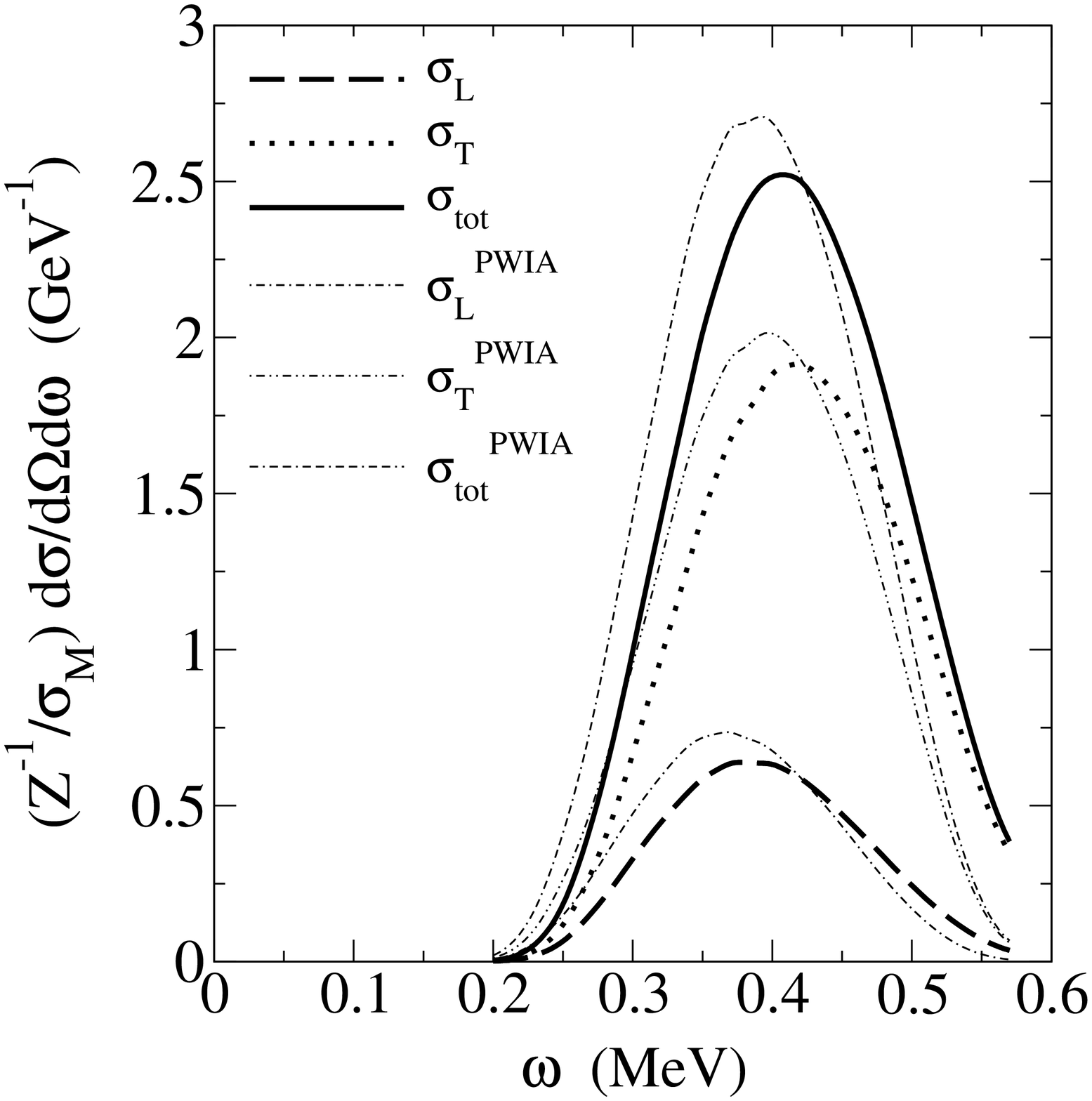}
\caption{ Ratio of EMAr and PWIA longitudinal, transverse and total cross
sections to Z times the Mott cross section for 800 MeV electron
scattering from $^{208}Pb$ at q=900 MeV/c.
}\label{fig:sigma.RUN2.Pb}
\end{figure}

\begin{figure}[h]
\includegraphics[width=10cm,bb= 0 0 600 600,clip]{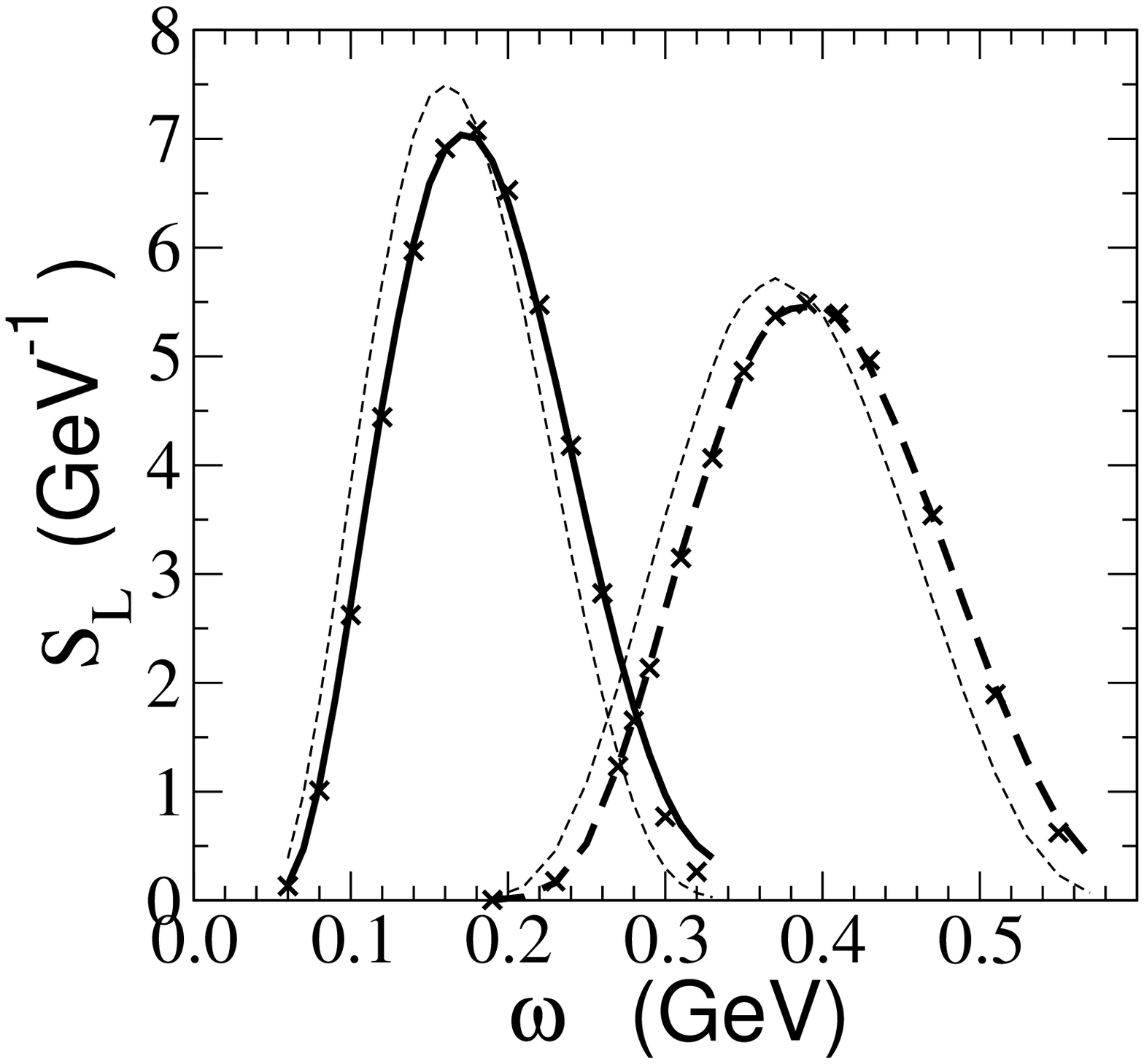}
\caption{ EMAr longitudinal response functions divided by Z and
the nucleon form factor for $^{208}$Pb at E$_i$ = 500 Mev and
$q=550$ MeV/c (solid line)
and at E$_i$ = 800 Mev and $q=900 MeV/c$ (dashed line).
The corresponding PWIA response functions without Coulomb effects
included are shown by the light dashed lines.
Fits of the
response functions using Eq.~(\ref{eq:EMAfit}) are shown by the $\times$
symbols and the parameters of the fits are given in Table~\ref{table:EMAfits}. }
\label{fig:RL.RUN1.RUN2.Pb}
\end{figure}

  Figure~\ref{fig:sigma.RUN1.Pb} shows cross sections for $^{208}Pb$
    at 500 MeV electron energy.  The momentum transfer is held fixed at 550 MeV/c
    and therefore the scattering angle varies with energy loss $\omega$ from about $70^{\small o}$
    to about $100^{\small o}$.  The figure shows the plane-wave (PWIA) cross sections as light
    lines and the EMAr results for Coulomb distorted cross sections as heavy lines.  Generally the
    Coulomb corrections shift the peaks to larger energy loss.  In our calculations,
    the average binding energy of 8 MeV was used for all shells.
    Figure ~\ref{fig:sigma.RUN2.Pb} shows similar cross sections at
    800 MeV electron energy with the momentum transfer held fixed at 900 MeV/c.
    In this case the scattering angle varies from about $70^{\small o}$ to about $105^{\small o}$.
    At the higher momentum transfer, the longitudinal cross section is seen to be
    a small fraction of the total cross section even without the pionic contributions.
Longitudinal response functions for $^{208}$Pb are shown in Figure~\ref{fig:RL.RUN1.RUN2.Pb}.
In this figure, the PWIA response functions shown obey the Coulomb sum rule in
the form
\begin{equation}
\int d \omega \frac{S_L^{PWIA}(\omega,q)}{ 1 + \frac{\omega^2}{4M(M+\omega}} = 1,
\end{equation}
where the kinematical factor in the denominator cancels the
kinematical factors due to wave functions normalizations and
currents.
The correction is modest: the denominator factor is about
1.01 at $\omega = 0.2$ GeV and 1.03 at $\omega =0.4$ GeV.

Note that the contributions owing to correlations in the nuclear wave functions
are omitted in our calculations.  If they are small at the
q values shown, the Coulomb sum rule should be satisfied approximately.
Our wave functions are approximate and final-state interactions
of the knocked-out nucleon have been omitted. Cross sections
presented in this work are not expected to be very close to experimental
results, however, the nuclear model used is expected to be adequate for
testing the accuracy with which Coulombic effects can be removed.

We have fit the EMAr longitudinal response functions to
the EMA form as in Eq.~(\ref{eq:EMAfit})
using the same value of $q_{eff}$ in the prefactor of Eq.~(\ref{eq:RL_EMAr})
as in the PWIA response function.
The fits of the EMAr response function based on all shells yield
    similar values for $A$ and a little smaller values for $\delta$k compared with fits of the
 1s shell response function.
The fit parameters are summarized in Table~\ref{table:EMAfits}.
\begin{table}
\caption{EMA fit parameters for the EMAr response functions
at energy $E$ in GeV and momentum transfer $q$ in GeV/c.
The value of $\delta$k is in MeV.
} \label{table:EMAfits}
\begin{tabular}{|ccccccc|}
\hline
Nucleus     &~~$E$~~&~~$q$~~ &~~$\delta$k~~&~~$f_{EMA}$~~&~$A^{EMAr}$~&~$A^{DWBA}$~  \\
\hline
$^{208}Pb$~~&~~0.5~~&~~0.55~~&  -21.0 &  0.82  &  0.98  & $\approx$1.00  \\
$^{208}Pb$~~&~~0.8~~&~~0.90~~&  -19.5 &  0.76  &  0.985 & $\approx$1.00  \\
$^{58}Fe$~~ &~~0.5~~&~~0.55~~&  -8.8 &  0.71  &  0.99  & $\approx$1.00 \\
$^{58}Fe$~~ &~~0.8~~&~~0.90~~&  -9.5 &  0.77  &  1.00  & $\approx$1.00 \\
\hline
\end{tabular}
\end{table}
Accounting for the nonlocality of the photon propagator as in
the results based on the $^{208}Pb$ 1s-shell DWBA response, the
$A^{EMAr}$ factors are renormalized by the
factor 1.015 to estimate factors $A^{DWBA}$ for a full DWBA analysis
that includes all shells.  Our results support the use of the
EMA fits of experimental data as in Eq.~(\ref{eq:EMAfit}) using $A=1$.
The $f_{EMA}$ factors are a little smaller when all shells are included. That is understandable
because higher shells include ones with wave functions that
vanish at $r=0$.  For those shells,
the distortion effects in the electron waves contribute at radii away
from $r=0$ where the Coulomb potential is weaker.  The results for the 1s shell show that
both EMAr and DWBA yield the same value of $\delta$k. The momentum shifts should be equal
also for response functions based on the sum over all shells.

\begin{figure}[h]
\includegraphics[width=10cm,bb= 0 0 600 600,clip]{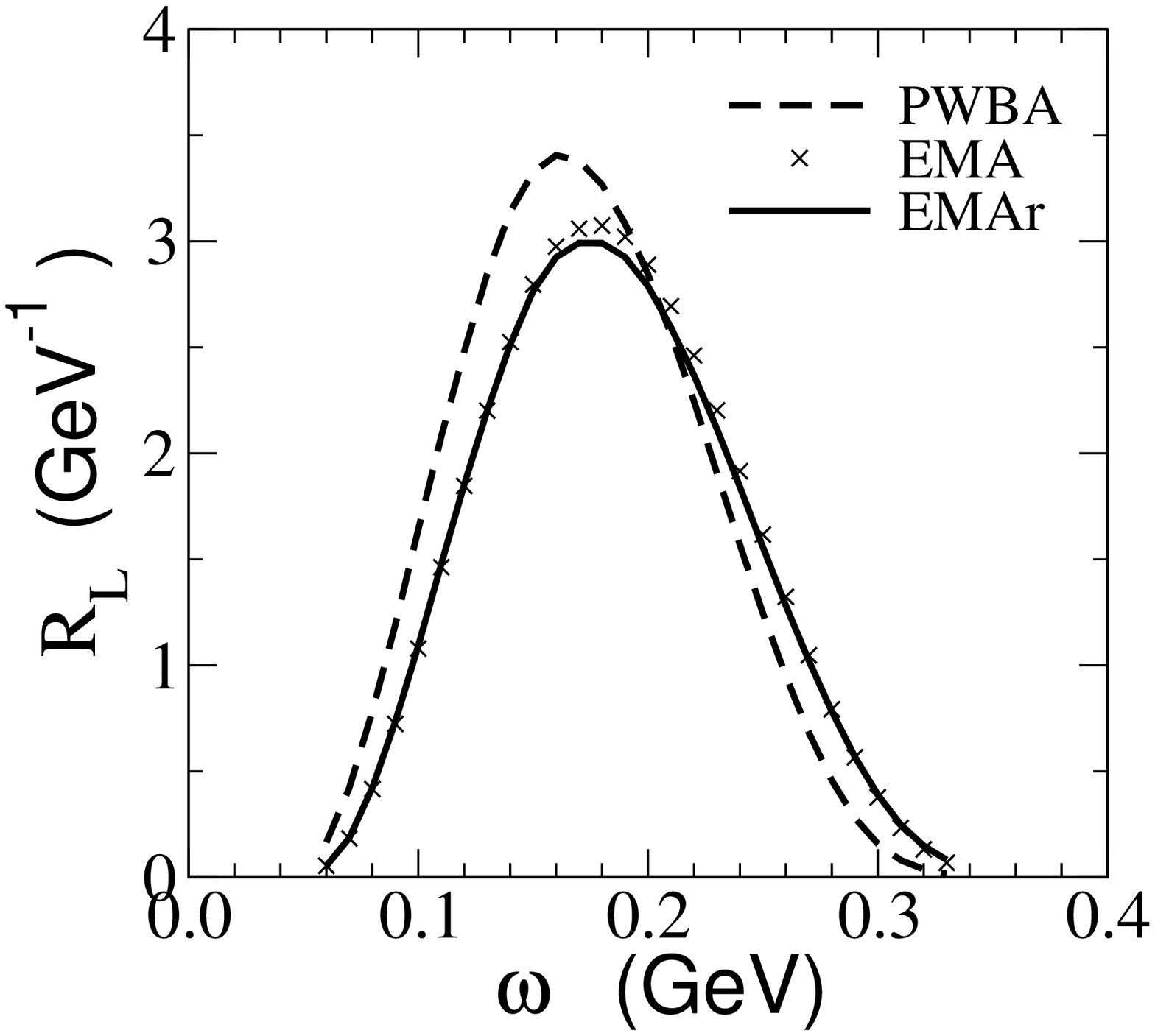}
\caption{ EMAr longitudinal response functions divided by Z 
for $^{208}$Pb at E$_i$ = 500 Mev and
$q=550$ MeV/c (solid line). 
The corresponding PWIA response function without Coulomb effects
included is shown by the dashed line and the EMA calculation using
$f_{EMA}=0.8$ and $A=1.0$ is shown by the $\times$ symbols. }
\label{fig:RL.RUN1.Pb.05.16.08} 
\end{figure}
\begin{figure}[h]
\includegraphics[width=10cm,bb= 0 0 600 600,clip]{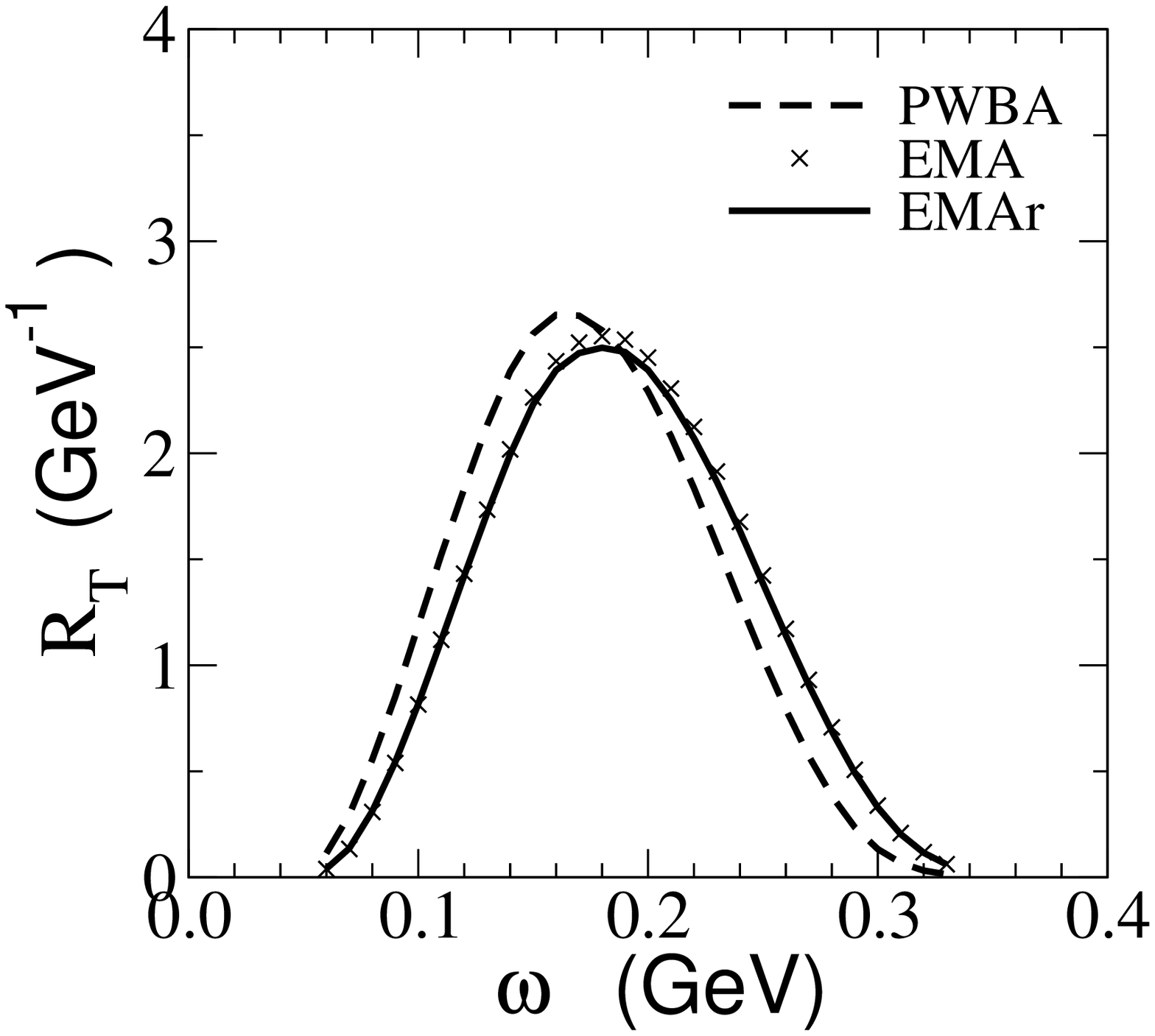}
\caption{ EMAr transverse response functions divided by Z 
for $^{208}$Pb at E$_i$ = 500 Mev and
$q=550$ MeV/c (solid line). 
The corresponding PWIA response function without Coulomb effects
included is shown by the dashed line and the EMA calculation using
$f_{EMA}=0.8$ and $A=1.0$ is shown by the $\times$ symbols. }
\label{fig:RT.RUN1.Pb.05.16.08}
\end{figure}

    Figures~\ref{fig:RL.RUN1.Pb.05.16.08} and \ref{fig:RT.RUN1.Pb.05.16.08}
show the longitudinal and transverse response functions $R_L$ and $R_T$
that do not have form factors divided out for Pb at q=550 MeV/c.  Results are shown for the
PWIA, EMA and EMAr calculations, where the EMA results are not a fit 
but rather are a straightforward calculation using $f_{EMA} = 0.8$ and $A=1.0$.
The Coulomb effects of the EMAr analysis are well approximated by the
EMA calculation. As has been discussed, the overall magnitude of the EMA
response is higher than the EMAr response 
by about 2\% because the $A$ parameter has not been used.  
  
\begin{figure}[h]
\includegraphics[width=10cm,bb= 0 0 612 792,clip]{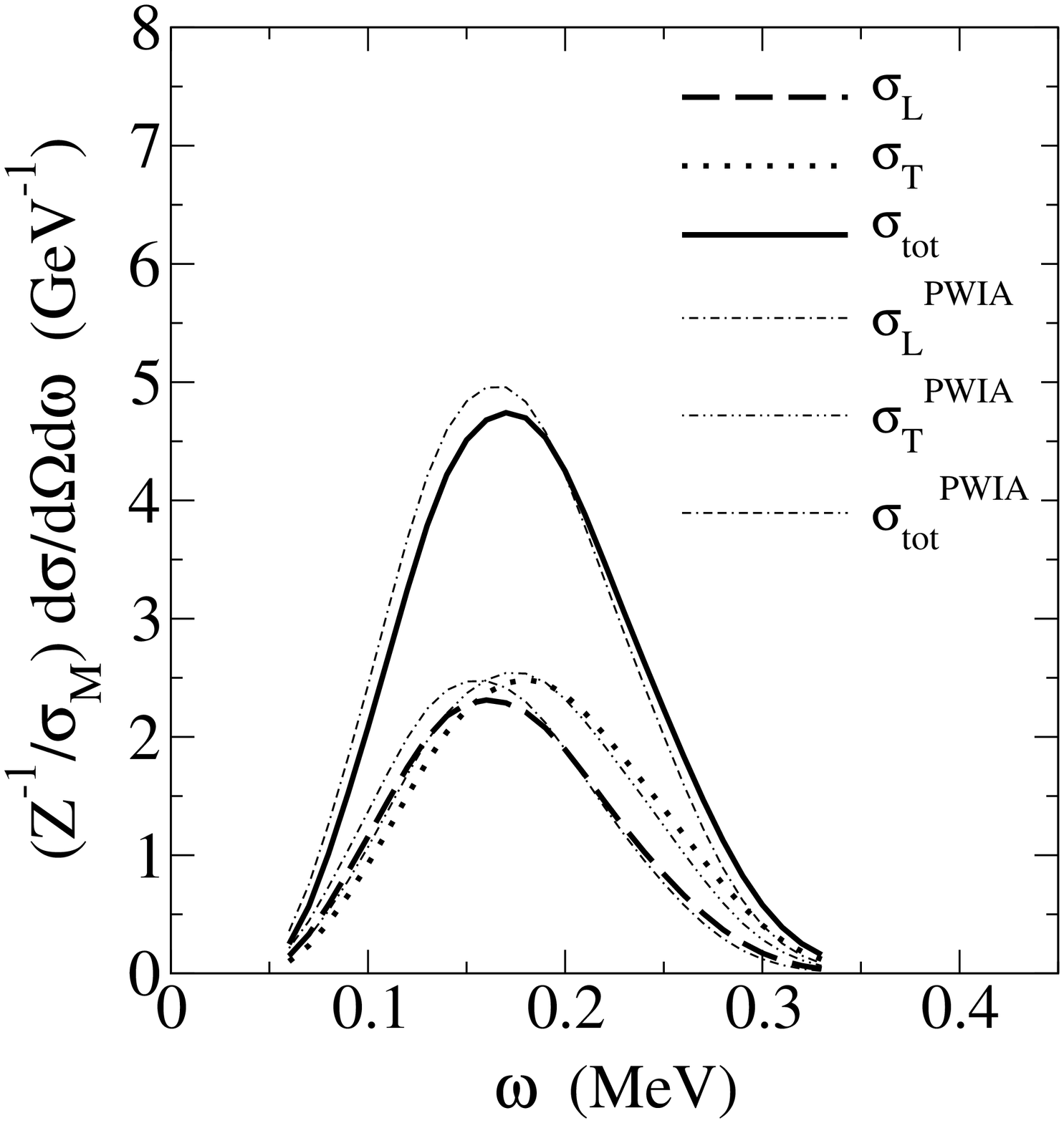}
\caption{ Ratio of EMAr and PWIA longitudinal, transverse and total cross
sections to Z times the Mott cross section for 500 MeV electron
scattering from $^{56}Fe$ at q=550 MeV/c.
}\label{fig:sigma.RUN1.Fe}
\end{figure}

\begin{figure}[h]
\includegraphics[width=10cm,bb= 0 0 612 792,clip]{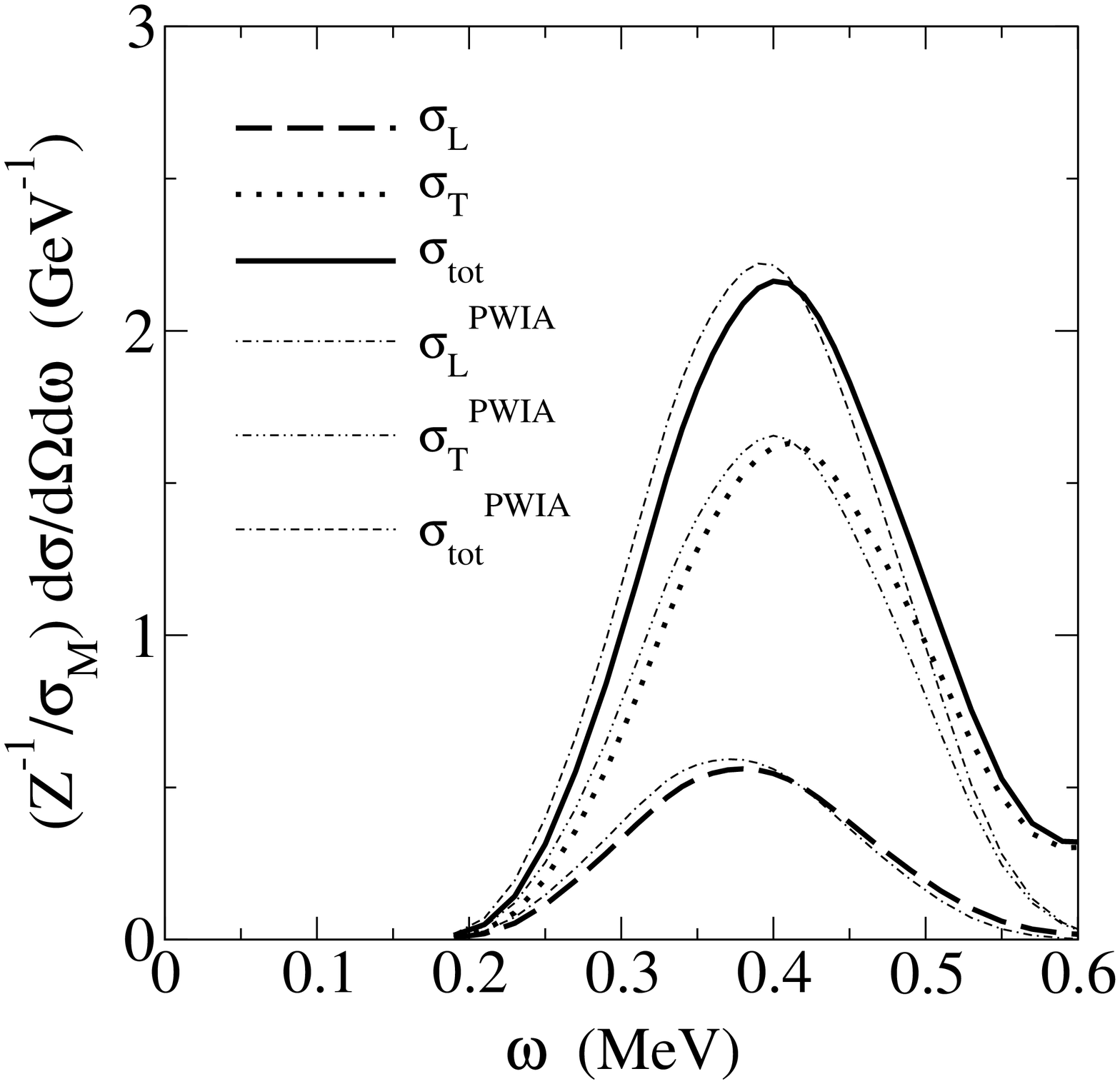}
\caption{ Ratio of EMAr and PWIA longitudinal, transverse and total cross
sections to Z times the Mott cross sections for 800 MeV electron
scattering from $^{56}Fe$ at q=900 MeV/c.
}\label{fig:sigma.RUN2.Fe}
\end{figure}

\begin{figure}[h]
\includegraphics[width=10cm,bb= 0 0 600 600,clip]{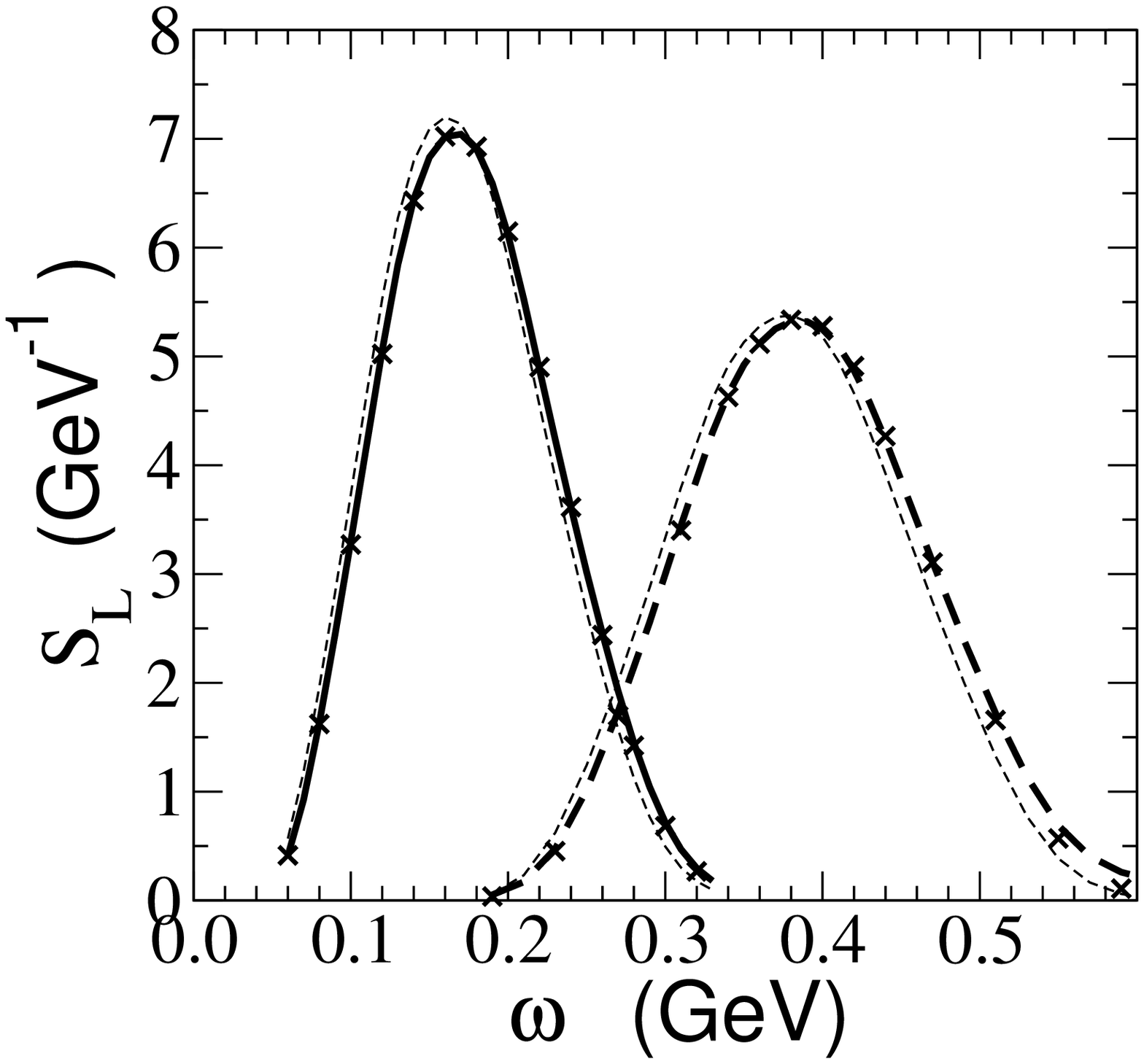}
\caption{ EMAr longitudinal response functions divided by Z
and the nucleon form factor for $^{56}$Fe at E$_i$ = 500 Mev and
$q=550$ MeV/c (solid line)
and at E$_i$ = 800 Mev and $q=900 MeV/c$ (dashed line).
The corresponding PWIA response functions without Coulomb effects
included are shown by the light dashed lines.  Fits of the
response functions using Eq.~(\ref{eq:EMAfit}) are shown by the $\times$
symbols and values of the fitting parameters are given
in Table~\ref{table:EMAfits}.
}\label{fig:RL.RUN1.RUN2.Fe}
\end{figure}

    Figure~\ref{fig:sigma.RUN1.Fe} shows cross sections for $^{56}Fe$
    at 500 MeV electron energy and q= 550 MeV/c
and Figure~\ref{fig:sigma.RUN2.Fe} shows cross sections at 800 MeV
electron energy and q = 900 MeV/c.  Coulomb effects are somewhat smaller
for the $^{56}Fe$ nucleus because the Coulomb potential is smaller.
Response functions for the $^{56}Fe$ target are shown in
Figure~\ref{fig:RL.RUN1.RUN2.Fe}. Fits of the response functions to the
EMA form of Eq.~(\ref{eq:EMAfit}) yield the fitting parameters
shown in Table~\ref{table:EMAfits}.  The shifts are given in this case by $f_{EMA} =$ 0.71 (500 MeV) and 0.77 (800 MeV).


\section{Conclusions}

    In this work we have tested some assumptions that have been
used in the analysis of experimental data for quasielastic scattering from
nuclei.  The main focus is to use a known nuclear
model (in this case the shell model) in order to test how well the Coulomb
corrections can be removed from DWBA cross sections using the effective-momentum
approximation (EMA).  The goal is to extract PWIA
response functions from the DWBA cross sections. It is assumed that
the Coulomb corrections are not much affected by the nuclear
model used.

   At the electron beam energies considered in the work, namely 500 MeV and higher,
the Coulomb effects in quasielastic scattering from nuclei can be described
accurately using the eikonal distorted waves that include higher-order corrections.
The eikonal analysis has simplifying features because one can isolate the
phases that cause shifts of the electron momenta, the focusing
factors and the spin phases that affect the L/T separation.  We have
used the analytical phases up to order $1/k^2$ in the eikonal
expansion that were developed in Ref.~\cite{Tjon&Wallace06}.  As one
check on the numerics, the eikonal
phases were computed two ways: by direct numerical integration of
the defining equations and by use of the analytical formulas.
Both give the same results. For the cases considered in this
work, the eikonal wave functions provide very well converged results.
As a check of the three-dimensional integration used in the EMAr
analysis, the PWIA results were computed
two ways: using analytical Fourier transforms of the
nucleon's bound state wave functions and by three-dimensional
numerical integration.  The latter calculations are the same
as those for the EMAr amplitude except that the Coulomb effects are
omitted.  With suitable integration grids the results are
essentially the same at an accuracy better than 1\% near the peak
of response functions and errors at larger $\omega$ can be
1\% or 2\% of the peak value of the response function. Generally
the errors in numerical results are insignificant in
the plots.

Full DWBA computations are extremely time consuming.
An approximation called EMAr is used to simplify the analysis.
The EMAr analysis evaluates the full r-dependence of the
eikonal distorted waves but approximates the hard-photon
propagator and the form factor in the nucleon current by evaluating them
at the effective momentum, $q_{eff}$.  Tests of the EMAr against
the full DWBA analysis were carried out for the response function of the 1s
shell of $^{208}$Pb.  Those tests showed that the EMAr produces close agreement with the
DWBA.  Moreover, the assumption that one should remove the nucleon
form factor (which is integrated over in the DWBA analysis) by evaluating
it at $q_{eff}$ was found to be justified with better than 1\% accuracy.
This should be compared with large
differences in cross sections when the form factor is evaluated
at $q$, the momentum transfer of the electron, instead of $q_{eff}$.
We find clear evidence that the form factor should be evaluated at the
effective momentum when it is divided out of experimental
cross sections in order to check the Coulomb sum rule.

The analysis of Bates experimental data in Ref.~\cite{Williamson97}
uses the form factor at $q$ rather $q_{eff}$ for a $^{40}Ca$ nucleus.
Results for the Coulomb sum rule are about 0.8-0.9 compared
with the expectation of unity.
If $q_{eff}$ were used in the analysis, the Bates results
for the Coulomb sum rule
would be increased by about 5\%, thus making them closer to unity.
The analysis of Saclay experimental data in
Refs.~\cite{Meziani84,Meziani85} uses form factors at $q_{eff}$.
Significantly lower values for the Coulomb sum rule are found
based on the Saclay analysis.  The differences between the
Bates and Saclay results are much larger than can be attributed to
Coulomb corrections.

We find that the spin phases in electron wave functions produce very
small effects at energies of 500 MeV or higher.  The helicity
matrix elements that involve the spin phases are very close to
those of a PWIA analysis for quasi-elastic scattering.
Consequently, the Rosenbluth separation extracts response functions
$R_L^{expt}$ and $R_T^{expt}$ that are accurate
in the sense that they correspond very closely to
the distorted wave matrix elements of the longitudinal and transverse parts of the currents.

The effects of the distorted waves on the longitudinal response function
are twofold: 1.) for electron scattering they shift the peak of the response functions towards larger values of the
energy loss, $\omega$, and 2.) they distort the shapes of the response
functions, more so for the inner shells than the outer ones.
However, reasonably accurate fits of the distorted response functions
can be obtained using the EMA fitting procedure of
Eq.~(\ref{eq:EMAfit}).
The momentum shift parameter $\delta k$ is found to be given by
$f_{EMA} \approx 0.80$,
for both the $^{208}Pb$ and $^{56}Fe$ nuclei, i.e., $\delta k \approx 0.80 V_c(0)$,
where $V_c(r)$ is the Coulomb potential.  More precise values are given in Table~\ref{table:EMAfits}.
The normalization parameter
$A$ is equal to 1.00 within
one or two percent.   The uncertainty arises because the normalization
for the sum over shells has
been calculated based on the full DWBA for the 1s-shell of $^{208}Pb$
and because the shape of the
distorted response function differs a little from the shape of the
PWIA response function for $\omega$ significantly away from the peak.
Therefore fits to the PWIA shape cannot reproduce the response precisely.
Note that the good agreement of $f_{EMA}$ and $A$ for $^{208}$Pb and $^{56}$Fe
demonstrates that the Coulomb corrections do not depend significantly
on the nuclear model.  Note also that the analysis of experimental data
using a fit as in Eq.~(\ref{eq:EMAfit}) tends to give more accurate results for
$R_L^{(PWIA)}$ at the peak of the response because that is controlled by $\delta k$
and less accurate results away from the peak because of the distortion of the shape.

   Estimates of longitudinal, transverse and total cross sections
have been calculated using shell model wave functions for $^{208}Pb$ and
$^{56}Fe$ at $q$ = 0.55 GeV/c and $q$ = 0.8 GeV/c.  These kinematical
conditions match the ones used in a recent experiment at Jlab. Because
final-state interactions, correlations and pion production have been omitted,
the calculated cross sections may differ significantly from
experimental cross sections.  Nevertheless the Coulomb corrections
should be reliable at the level of a few percent.

Coulomb corrections are notoriously difficult to
calculate and our calculations refute
claims that may be found in the literature. For example,
Ref.~\cite{Kim05} claims that the EMA procedure is not
accurate for the longitudinal response
at 485 electron energy and $60^{o}$ scattering angle for a $^{208}Pb$ target.
The basis for the claim is that significant differences are found between
EMA results and results
based on an ad-hoc DWBA analysis that has been used extensively.
We find that the EMA with appropriate parameters can describe
the 1s-shell DWBA or all-shells EMAr results very well at
essentially the same kinematics.
We wish to emphasize that all of our
numerics are under good control and various
consistency checks have been made that give
confidence in the results reported herein.

\end{document}